\documentclass[11pt,a4paper]{article} %11pt?
\usepackage{jheppub,multirow}
\usepackage{amsthm}
\usepackage{xcolor}
\usepackage{float}
\usepackage{url}
\usepackage{microtype}
\usepackage{subcaption}
\usepackage{empheq}
\usepackage{booktabs}  
\usepackage[makeroom]{cancel}
\usepackage{dsfont} %Nice \1
\usepackage{ytableau} %Make Young tableaux
\usepackage{setspace}%Set line spacing to 1.15
\usepackage[skins]{tcolorbox}

\newcommand{\beq}{\begin{equation}}
\newcommand{\eeq}{\end{equation}}

\newcommand{\bR}{\mathbb{R}}

\newcommand{\fq}{\mathfrak{q}}

\newcommand{\cN}{\mathcal{N}}

\newcommand{\cT}{\mathcal{T}}

\newcommand{\E}{{\epsilon}}

\newcommand{\sh}{\mbox{sh}}
\newcommand{\ch}{\mbox{ch}}
\def\ie{\begin{equation}\begin{aligned}}
\def\fe{\end{aligned}\end{equation}}
\def\xTB{$\times$} % inside table
\newcommand*\xbar[1]{%
	\hbox{%
		\vbox{%
			\hrule height 0.5pt % The actual bar
			\kern0.3ex%         % Distance between bar and symbol
			\hbox{%
				\kern-0.1em%      % Shortening on the left side
				\ensuremath{#1}%
				\kern-0.05em%      % Shortening on the right side
			}%
		}
	}
}

\newtheoremstyle{fullit}
{\topsep}      % ABOVESPACE
{\topsep}      % BELOWSPACE
{\normalfont}  % BODYFONT % \itshape
{0pt}          % INDENT (empty value is the same as 0pt)
{\itshape}     % HEADFONT
{.\ }          % HEADPUNCT
{0pt}          % HEADSPACE. `plain` default: {5pt plus 1pt minus 1pt}
{\thmname{#1} \thmnumber{#2}}             % CUSTOM-HEAD-SPEC

\theoremstyle{definition}

\theoremstyle{fullit}

\DeclareCaptionSubType*[alph]{figure}
\title{Quantum Geometry and $\theta$-Angle in Five-Dimensional Super Yang-Mills}
\author{Nathan Haouzi}
\affiliation{Simons Center for Geometry and Physics, State University of New York,
	Stony Brook, NY 11794}
\emailAdd{nhaouzi@scgp.stonybrook.edu}

\abstract{Five-dimensional $Sp(N)$ supersymmetric Yang-Mills admits a $\mathbb{Z}_2$ version of a theta angle $\theta$. In this note, we derive a double quantization of the Seiberg-Witten geometry of $\mathcal{N}=1$  $Sp(1)$ gauge theory at $\theta=\pi$, on the manifold $S^1\times\mathbb{R}^4$. Crucially, $\mathbb{R}^4$ is placed on the $\Omega$-background, which provides the two parameters to quantize the geometry. Physically, we are counting instantons  in the presence of a 1/2-BPS fundamental Wilson loop, both of which are wrapping $S^1$. Mathematically, this amounts to proving the regularity of a $qq$-character for the spin-1/2 representation of the quantum affine algebra $U_q(\widehat{A_1})$, with a certain twist due to the $\theta$-angle.  We motivate these results from two distinct string theory pictures. First, in a $(p,q)$-web setup in type IIB, where the loop is characterized by a D3 brane. Second, in a type I' string setup, where the loop is characterized by a D4 brane subject to an orientifold projection. We comment on the generalizations to the higher rank case $Sp(N)$ when $N>1$, and the $SU(N)$ theory at Chern-Simons level $\kappa$ when $N>2$.}
\begin{document}
	\maketitle
	\setlength{\parindent}{0pt}
	\clearpage
	
	%%%%%%%%%%%%%%%%%%%%%%%%%%%%%%%%%%%%%%%%%%%%%%%
	%%%%%%%%%%%%%%%%%%%%%%%%%%%%%%%%%%%%%%%%%%%%%%%
	%%%%%%%%%%%%%%%%%%%%%%%%%%%%%%%%%%%%%%%%%%%%%%%
	%%%%%%%%%%%%%%%%%%%%%%%%%%%%%%%%%%%%%%%%%%%%%%%
	%%%%%%%%%%%%%%%%%%%%%%%%%%%%%%%%%%%%%%%%%%%%%%%
	%%%%%%%%%%%%%%%%%%%%%%%%%%%%%%%%%%%%%%%%%%%%%%%
	%%%%%%%%%%%%%%%%%%%%%%%%%%%%%%%%%%%%%%%%%%%%%%%
	%%%%%%%%%%%%%%%%%%%%%%%%%%%%%%%%%%%%%%%%%%%%%%%

\newpage
	
\section{Introduction}
	
\subsection{$SU(2)_{\pi}$ Gauge Theory and $\widetilde{E_1}$ Fixed Point}

The dimension ``five" holds a special place in the landscape of supersymmetric gauge theories, since the theories are non-renormalizable. Still,  there is by now a large amount of evidence that strongly-interacting supersymmetric theories exist as UV fixed points in 5d. The gauge theory is then understood as a relevant deformation in the IR, where the gauge coupling is set by such a deformation parameter.
	
Perhaps the most famous example is the 5d $Sp(1)=SU(2)$ gauge theory, with $0 \leq N_f \leq 7$  fundamental hypermultiplets.  The action has a $SO(2 N_f) \times U(1)$ global symmetry, with $SO(2 N_f)$ the flavor symmetry, and $U(1)$ a topological symmetry associated to the conserved current $J=\ast Tr(F\wedge F)$.  It was argued by Seiberg \cite{Seiberg:1996bd} that such a theory has a UV fixed point where the symmetry group $SO(2 N_f) \times U(1)$ gets enhanced to $E_{N_f+1}$. With the recent success of localization methods, this symmetry enhancement was checked explicitly through a computation of the $Sp(1)$ superconformal index \cite{Kim:2012gu}, and an index for 1/2-BPS ray operators \cite{Chang:2016iji}.\\

\begin{figure}[h!]
	%\begin{center}
	\emph{}
	%\hspace{-20ex}
	\centering
	\includegraphics[trim={0 0 0 3cm},clip,width=0.8\textwidth]{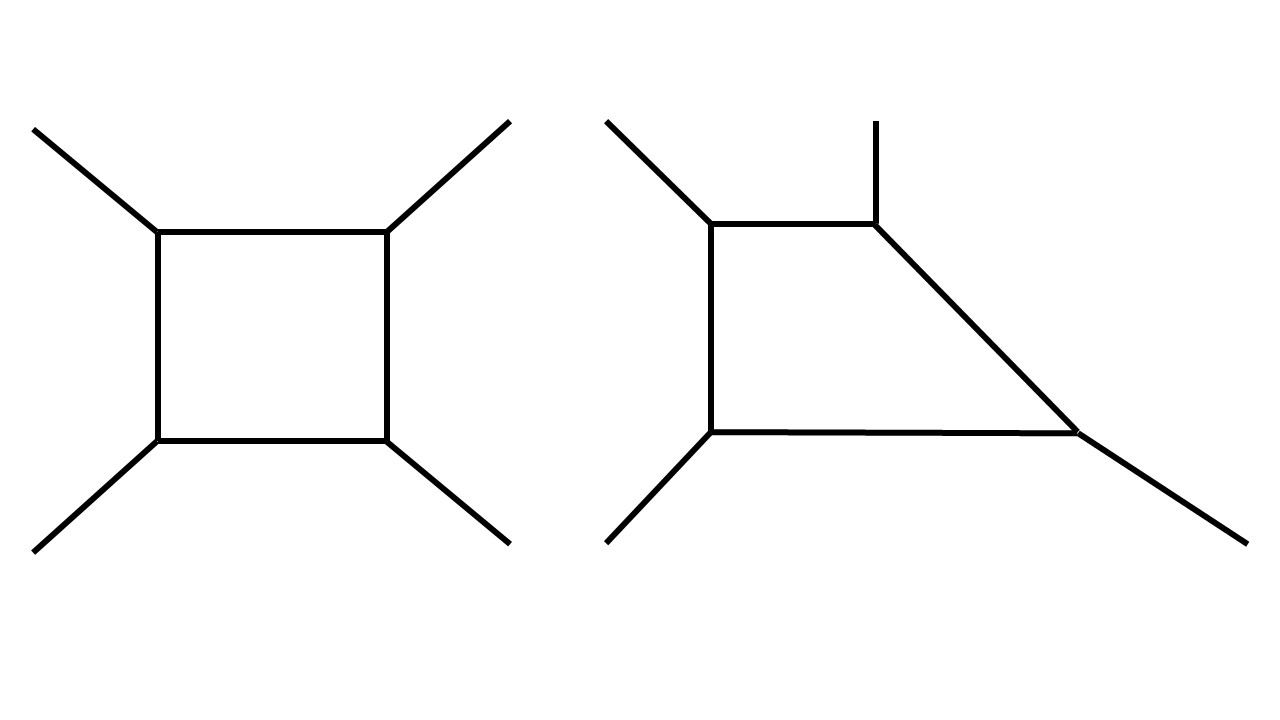}
	\vspace{-12pt}
	\caption{On the left, the $SU(2)$ gauge theory at $\theta=0$, which flows to the fixed point $E_1$ when we collapse the web. On the right, the $SU(2)$ gauge theory at $\theta=\pi$, which flows to the fixed point $\widetilde{E_1}$ when we collapse the web.} 
	\label{fig:E1andE1tilde}
	%\end{center}
\end{figure} 

In the absence of fundamental flavors, $N_f= 0$, it was argued \cite{Morrison:1996xf} that there should exist a second fixed point theory, the so-called $\widetilde{E_1}$ theory, where the $U(1)$ topological symmetry does not get enhanced.  To justify its existence, one can proceed as follows: consider the $E_2=SU(2)\times U(1)$ fixed point theory. There exist two relevant deformation parameters: a flavor mass $m$, associated to the $U(1)$, and a bare inverse gauge coupling $1/g_0^2$, with the combination $m_0\equiv 1/g_0^2-2m$ associated to the $SU(2)$. The locus $\{m, m_0=0\}$ is the fixed point theory $E_2$. When $m=0$ and $m_0>0$, we flow to a free theory, with global symmetry $SO(2)\times U(1)$. When $m>0$ and $m_0=0$, we flow to the $E_1$ theory, with global symmetry $E_1=SU(2)$. Interestingly, when $m<0$ and $m_0>0$, there is a singularity at $m_0 + 4m =0$, which suggests the existence of another fixed point: this is the $\widetilde{E_1}$ theory, with global symmetry $U(1)$. It follows that the theory has only one relevant deformation parameter, the combination $m_0 + 4m$.

In the rest of this note, we denote the associated gauge theory as $SU(2)_{\pi}$. The subscript notation was chosen to reflect the fact that in five dimensions, there exists a $\mathbb{Z}_2$ analog of the $\theta$ parameter familiar from four-dimensional physics \cite{Douglas:1996xp}. Indeed, in 4d, one has $\pi_3(SU(2))=\mathbb{Z}$, so there exist nontrivial classical gauge field configurations in Euclidean $\mathbb{R}^4$ with instanton number $k\in\mathbb{Z}$. Then, given a configuration with instanton number $k$, the path integral picks up  a phase $e^{i \theta k}$, and the shift $\theta\rightarrow\theta+2\pi$ is a symmetry of the theory, even at the quantum level. 
It so happens that $\pi_4(SU(2))=\mathbb{Z}_2$ (this nontrivial fourth homotopy group is also the reason for Witten's global anomaly in four dimensions \cite{Witten:1982fp}), so we can think of $\theta$ as a $\mathbb{Z}_2$-valued instanton in 5d.\\

\subsection{Non-Perturbative Schwinger-Dyson Equations}

Our main goal in this paper will be to derive a quantization of the $SU(2)_{\pi}$ Seiberg-Witten geometry. Equivalently, this result can be understood as a non-perturbative version of Schwinger-Dyson type identities for the theory \cite{Nekrasov:2015wsu}, as we shall now review. 

Consider the vacuum expectation value (vev) of some operator $Y$ in quantum field theory. Given the correlator $\langle Y \rangle$ defined by a path integral, the Schwinger-Dyson equations can be understood as constraints that must be satisfied by such a correlator. This comes about from demanding that the path integral remain invariant under a slight shift of the contour (provided that the measure is left invariant by such a shift).  
We are interested here in a particular contour modification that takes us from a given topological sector of $SU(2)_{\pi}$  to another distinct topological sector, related to the first by a large gauge transformation. In this case, the contour is discrete, and the transformation changes the instanton number of the theory. Then, our task is to construct an appropriate operator $Y$ to mediate such a change in instanton number, and the Schwinger-Dyson equations will impose constraints on the regularity of its vev. Put differently, we want to construct a  codimension-4 defect operator in $SU(2)_{\pi}$. Since the gauge theory is defined on the manifold $S^1\times \mathbb{R}^4$, the instantons are wrapping $S^1$, so a natural candidate for our problem is a supersymmetric Wilson loop operator wrapping that same circle.\\

A Wilson loop is formulated as the trace of a holonomy matrix, where a quark is parallel-transported along a closed curve in the manifold, and the trace is evaluated in some irreducible representation of the gauge group. Here, we will consider a 1/2-BPS loop wrapping $S^1$, located at the origin of $\mathbb{R}^4$, and the trace will be evaluated in the fundamental representation (or spin 1/2) of $SU(2)_{\pi}$. 
Denoting the quark mass as $M$, we will construct a loop operator vev as a function of $M$, which we call $\langle Y(M) \rangle$. This vev will turn out to  have generic poles in $M$, but a particular combination of $Y$-operator vevs will be regular in $M$; this is the content of the Schwinger-Dyson identity.

The identity will come about from counting instantons in the presence of the loop quark. Since instantons in five dimensions are particles, counting them amounts to computing the partition function of their quantum mechanics, which is an equivariant Witten index. One needs to treat carefully the contribution of coincident zero-size instantons, as the moduli space is singular there. In our case, the instantons are also coincident with the Wilson loop, so regularizing their contribution must be done with extra care. In general, the so-called ADHM \cite{Atiyah:1978ri} construction is a powerful way to resolve such singularities. In the case at hand, the instantons have a UV description as a $\cN=(0,4)$ ADHM gauged quantum mechanics \cite{Tong:2014cha}\footnote{Throughout this paper, when we talk about $\cN=(0,2)$ or $\cN=(0,4)$ supersymmetry in the context of the quantum mechanics, what we really mean is the reduction of two-dimensional $\cN=(0,2)$ or $\cN=(0,4)$ supersymmetry to one dimension.}. 

Crucially, the computation of the partition function relies on the fact that our five-dimensional manifold admits a two-parameter deformation known as the $\Omega$-background, which can be thought of as a weak $\cN=2$ supergravity background \cite{Nekrasov:2002qd,Nekrasov:2003rj}. Then, the Schwinger-Dyson identity will depend on the two parameters of the background. In the flat space limit, the identity becomes the Seiberg-Witten curve of the $SU(2)_{\pi}$ gauge theory. It is in that sense that  the Schwinger-Dyson identity in the $\Omega$-background should be thought of as a double quantization of Seiberg-Witten geometry.\\

There exists an important connection to representation theory, first pointed out in the absence of $\theta$-angle in \cite{Nekrasov:2015wsu}. Namely, the instanton partition function in the presence of a supersymmetric codimension-4 defect has been dubbed a $qq$-character of a finite dimensional irreducible representation of the quantum affine algebra $U_q(\widehat{A_1})$\footnote{The literature on the representation theory of quantum affine algebras is rich, and remains an active subject of research. For a description of finite dimensional representations, there are two popular presentations, one due to Jimbo \cite{jimbo}, and another due to Drinfeld \cite{Drinfeld:1987sy}. In our setting, it is the latter presentation that is relevant. In particular, see the works   \cite{Chari:1994pf, Chari:1994pd}. Characters of finite-dimensional representations of quantum affine algebras were constructed under the name $q$-characters \cite{Frenkel:qch}, in the context of studying (deformations of) $W$-algebras in two-dimensional conformal field theory. A deformed character depending on two parameters was  proposed in \cite{Shiraishi:1995rp,Awata:1995zk,Frenkel:1998} (for related  work on $t$-analogues of $q$-characters, see \cite{Nakajima:tanalog}).  This $qq$-character is precisely the one rediscovered in supersymmetric gauge theories by Nekrasov \cite{Nekrasov:2015wsu}, where ``$qq$" stands for the two parameters of the five-dimensional $\Omega$-background.  The fact that an object initially defined in the context of two-dimensional conformal field theory makes an appearance as an observable in five-dimensional supersymmetric gauge theory is an example of what is sometimes referred to as the BPS/CFT correspondence.}. In particular, the $SU(2)_{\pi}$ partition function can be presented as such a character, with a certain twist due to the nontrivial $\theta$-angle. Such a twist will be present in general for $SU(N)$ gauge theories at Chern-Simons level $\kappa$ when $N>2$, and we will derive the corresponding $qq$-character for that case as well.\\

Finally, as a byproduct of the Schwinger-Dyson identity, we will be able to compute the exact vev of a $SU(2)_{\pi}$ Wilson loop in spin 1/2 representation, including the instanton corrections.\\

\subsection{Organization of the Paper}

We will derive the quantum mechanics index and the associated quantized Seiberg-Witten geometry in two different ways.\\

In section 2, we treat the $SU(2)_{\pi}$ theory as a $U(2)$ theory with a Chern-Simons term (only $SU(N)$ with $N>2$ can admit a Chern-Simons term),  in the presence of a 1/2-BPS Wilson loop. We then freeze an overall $U(1)$ inside $U(2)$. This can be motivated from string theory. Indeed, it is known that the $SU(2)_{\pi}$ theory can be realized using a web of $(p,q)$-5 branes in type IIB  \cite{Aharony:1997bh}. In this picture, the loop defect is realized as a D3 brane inside the web \cite{Assel:2018rcw}.\\

In section 3, we compute instead the index using $Sp(N)_{\pi}$ instanton calculus, specialized to $N=1$. We find agreement with the index computed in section 2. We can again motivate the result from string theory; this time around, it is more natural to work in the type I' string, where the gauge theory is the effective theory on a D4 brane on top of an O8-orientifold plane. In that picture, the loop defect is realized as a codimension-4 D4 brane.\\

The formalism used in this paper allows several generalizations, which will be the subject of section 4. Namely, for the same amount of work, we will be able to derive the quantized geometry of 5d $SU(N)$ gauge theory, $N>2$, with Chern-Simons level $\kappa$. We will also comment on the quantized $Sp(N)_{\pi}$ geometry when $N>1$, and the addition of fundamental matter.\\

Before proceeding with our analysis, we deem it useful to first remind the reader of what has been done in the literature, and provide a (non-exhaustive) list of references:\\

When $\theta=0$, the study of 1/2-BPS line defects in 5d $\cN=1$ $SU(2)$ gauge theory has been an active topic of research  in the last few years. The investigation of the instanton moduli space was initiated in \cite{Tong:2014cha}, with an underlying type IIA string theory picture, and the partition function was computed as a Witten index in \cite{Kim:2016qqs}. That same partition function was interpreted as a set of non-perturbative Schwinger-Dyson equations in \cite{Nekrasov:2015wsu}. A $q$-CFT perspective related to a deformation of the Virasoro algebra was given in \cite{Kimura:2015rgi}, see also \cite{Mironov:2016yue} for a matrix model point of view. The line defects were interpreted as D2 branes in $(1,1)$  $A_1$ little string theory in \cite{Haouzi:2019jzk}. A T-dual perspective in type IIB was given in  \cite{Tong:2014yna}, which provided a UV description for a candidate holographic dual of $AdS_3\times S^3 \times S^3 \times S^1$. The precise analysis of the instanton moduli space in that background requires turning on $B$-fields, which was addressed in \cite{Nekrasov:2016gud}. 

Treating the $SU(2)$ gauge group as $Sp(1)$, an index for ray operators was defined based on a type I' string theory construction in \cite{Chang:2016iji}, building on the computation of the superconformal index in the absence of defect carried out in \cite{Kim:2012gu}. 

Yet another T-dual picture was given in terms of a type IIB $(p,q)$ web in \cite{Assel:2018rcw}, with the aim of studying how $S$-duality acts on Wilson loops. Realizing the $SU(2)_{\pi}$ geometry in that setup will prove to be particularly convenient for us, so we will follow that presentation in the next section. For an algebraic perspective, see \cite{Bourgine:2017jsi}.\\

When $\theta=\pi$, the physics of supersymmetric line defects in 5d $SU(2)_{\pi}$ has received little attention in comparison\footnote{For a recent notable exception, see \cite{Gaiotto:2015una} and the related investigation of Wilson loops in $SU(N)$ gauge theory at Chern-Simons level $N-\frac{N_f}{2}$, with $N_f$ fundamental flavors. There, a superconformal index is computed by localizing a Wilson loop on BPS configurations in the absence of the loop. This is not the approach we follow in this paper; instead, relying on a D-brane quantum mechanics, we argue that we should modify the definition of the index altogether when a Wilson loop is present. See the related discussion in \cite{Nekrasov:2015wsu,Kim:2016qqs,Chang:2016iji,Assel:2018rcw,Haouzi:2020yxy}.}. The partition function and superconformal index of the pure 5d $\cN=1$ $SU(2)_{\pi}$ gauge theory, in the absence of defects, were computed in \cite{Kim:2012gu,Bergman:2013ala}, and from the topological string in \cite{Iqbal:2012xm}.

A $(p,q)$ web analysis of the $SU(2)_{\pi}$ Seiberg-Witten geometry in type IIB string theory was carried out in \cite{Kim:2014nqa}. A realization in terms of orientifold O5 planes was proposed in \cite{Zafrir:2015ftn, Hayashi:2017btw}, and in terms of orientifold O7 planes in \cite{Bergman:2015dpa}.

\section{The $SU(2)_{\pi}$ quantized geometry, the type IIB way}

Our starting point is type IIB string theory in flat space. We compactify the $X_0$-direction on a circle $S^1(R)$ of radius $R$ and introduce the following branes:
\begin{table}[H]\centering
	\begin{tabular}{|l|cccccccccc|} \hline
		&  0 &  1 &  2 &  3 &  4 &  5 &  6 &  7 &  8 &  9  \\
		\hline\hline
		D5 &\xTB&\xTB&\xTB&\xTB&\xTB&\xTB&    &    &    &     \\
			\hline
		NS5 &\xTB&\xTB&\xTB&\xTB&\xTB&    &\xTB    &    &    &     \\
			\hline
		$(p,q)$-5 &\xTB&\xTB&\xTB&\xTB&\xTB&  $\phi$  & $\phi$  &    &    &     \\
		\hline\hline
		D1 &\xTB&   &   &   &   & \xTB   &    &    &    &    \\
		\hline
		F1 &\xTB&    &    &    &    & & \xTB & & &  \\
		\hline\hline
		D3 &\xTB&    &    &    &    &  &  &\xTB&\xTB&\xTB \\
		\hline
	\end{tabular}
	\caption{The $(p,q)$ web of branes. The direction $X_0$ is compact. The angle $\phi$ is defined by $\tan(\phi)=q/p$, and the various $(p,q)$ branes span the line $\cos(\phi) X_5 + \sin(\phi) X_6 =0$ in the $X_{5,6}$ plane.}
	\label{table:BranesSUN}
\end{table}

The particular brane setup we consider has two parallel D5 branes, which are horizontal segments along $X_5$; they support a 5d $\cN=1$ $SU(2)$ gauge theory on them \cite{Aharony:1997bh}. The theory is defined on the manifold $S^1(R)\times\mathbb{R}^4$, with the circle along $X_0$ and $\mathbb{R}^4$ along $X_{1, 2, 3, 4}$. We call the distance between the two D5 branes $2 a$, where the real scalar $a$ is the vev of the vector multiplet scalar. The F1 strings which stretch between the D5 branes realize W-bosons of mass $2 a$. We denote the bare Yang-Mills coupling as $t_0\equiv \frac{1}{g_0^2}$. The effective coupling of the theory is given by the length of the top D5 brane, which is determined by the geometry of the web to be $t_{eff}\equiv t_0 + a$. BPS instanton particles are realized as D1 strings located on the top D5 brane. The effective theory on $k$ such instanton strings is a one-dimensional quantum mechanics with gauge group $\widehat{G}=U(k)$. 

\begin{figure}[h!]
	%\begin{center}
	\emph{}
	%\hspace{-20ex}
	\centering
	\includegraphics[trim={0 0 0 3cm},clip,width=0.8\textwidth]{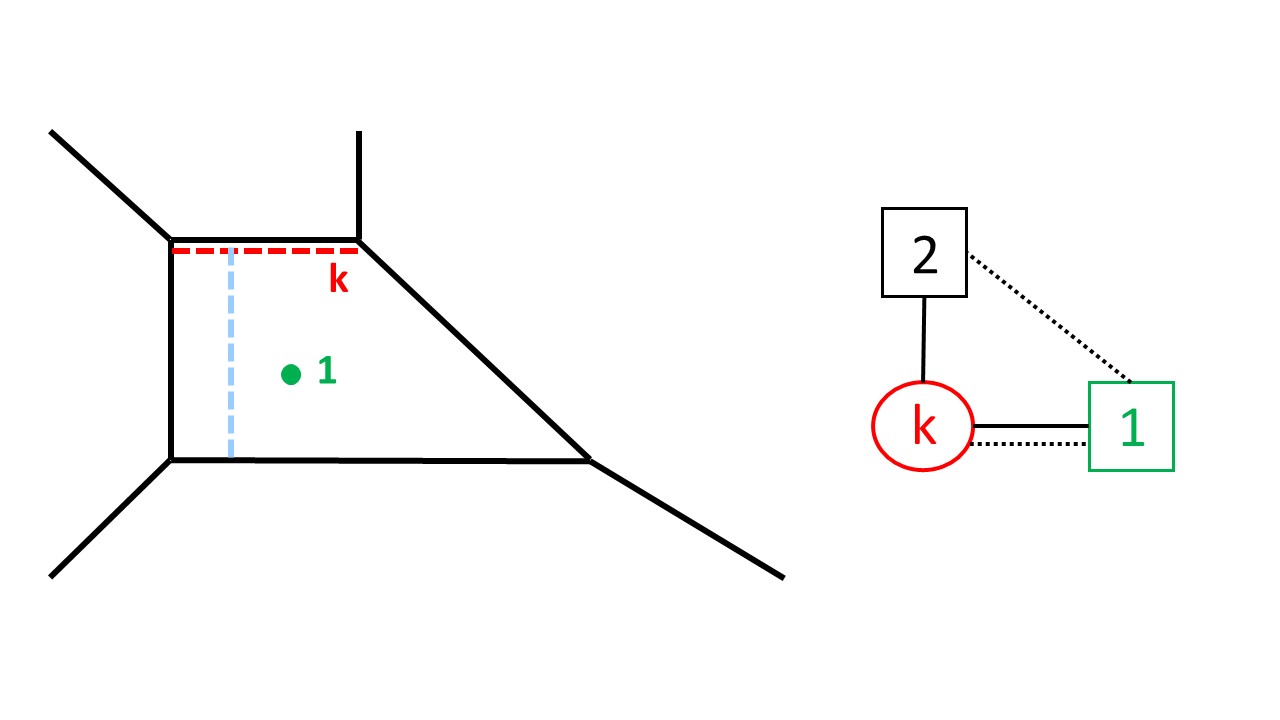}
	\vspace{-12pt}
	\caption{On the left, the $SU(2)_{\pi}$ theory living on the two horizontal (black) D5 branes. The D3 brane loop defect is the green dot, with gauge group $U(1)_{loop}$. The $k$ D1 instanton strings are the red horizontal segments, and the blue F1 string represents a $W$-boson of mass $2 a$. On the right, the $(0,4)$ quantum mechanics living on the D1 strings. It is a $U(k)$ gauge theory with a vector multiplet and adjoint hypermultiplet, represented by the red circle. There are two hypermultiplets in the bifundamental of $U(k)\times SU(2)$, represented by a single solid line. There is one twisted hypermultiplet and one Fermi multiplet (two complex fermions) in the bifundamental of $U(k)\times U(1)_{loop}$, represented by the mixed solid/dashed line. Finally, there are two Fermi multiplets  (a single complex fermion each) in the bifundamental of $SU(2)\times U(1)_{loop}$, represented by the one dashed line.} 
	\label{fig:E1tildeBPS}
	%\end{center}
\end{figure}

The web \ref{fig:E1tildeBPS} showcases a particular vacuum of the gauge theory on the Coulomb branch, along with BPS massive particles (W-boson and instantons). The $\widetilde{E_1}$ fixed point is reached when $t_{eff}, a \rightarrow 0$, which is when the face of the web ``collapses" to zero area.\\

Our main interest lies in introducing 1/2-BPS loop defects, which in the brane setup can be realized as semi-infinite F1/D1 strings \cite{Maldacena:1998im, Rey:1998ik}, but also as D3 branes to allow for loops in more general representations of the gauge group \cite{Drukker:2005kx,Gomis:2006sb,Yamaguchi:2006tq}. 

For our purposes, it will be enough to consider a single D3 brane\footnote{The  number of D3 branes determines which representation of $SU(2)$ the Wilson loop transforms in \cite{Gomis:2006sb}. $N$-fold tensor product representations can be constructed by considering $N$ D3 branes in the web. Since only the fundamental representation (spin 1/2) is relevant to make contact with Seiberg-Witten geometry, we only need to consider $N=1$ in this paper.} wrapping the $X_{0,7,8,9}$ directions and sitting at $X_{1, 2, 3, 4}=0$, see  table  \ref{table:BranesSUN}. The web is drawn in the $X_{5,6}$ plane, so the  D3 brane is a point on it. The total configuration of branes preserves four supercharges, and supersymmetry is broken to a (reduction from 2d to) 1d $\cN=(0,4)$ subalgebra.  The supermultiplets present in the D1 brane quantum mechanics are summarized in figure \ref{fig:E1tildeBPS}. There are also $J$ and $E$ potential terms required by $(0,4)$ supersymmetry, but their precise form is not needed in what follows; a good review can be found in \cite{Tong:2014yna,Putrov:2015jpa}.\\

A 4d $\cN=4$ $U(1)_{loop}$ theory lives on the D3 brane, and couples to the 5d theory along the loop in the $X_0$ direction. We treat the 4d theory as non-dynamical, so that the D3 brane can be interpreted as a one-dimensional 1/2-BPS loop defect in the 5d theory. The  modes of the 1d fields localized on the loop are two complex fermions $ \chi_{i=1, 2}$, transforming in the bifundamental of  $SU(2)\times U(1)_{loop}$. They couple to the bulk through the following term in the action:
\begin{equation}
\label{5d1ddfermion}
S^{5d/1d}=\int dt\; \chi_{i}^\dagger\, \left( \delta_{ij} \, \partial_t - i\,A^{[5d]}_{t, ij}   + \Phi^{[5d]}_{ij}   - \delta_{ij}\, M \right)\, \chi_{j} \; .
\end{equation}
Above, $A^{[5d]}_t$ and $\Phi^{[5d]}$ are the pullback of the 5d gauge field and the adjoint scalar of the vector multiplet, respectively. $i$ and $j$ are indices for the fundamental representation of $SU(2)$.  The parameter $M$ is the eigenvalue of the background (nondynamical)  $U(1)_{loop}$ gauge field. It sets the energy scale for the excitation of the fermions. The variable $t$ is periodic, with period $R/(2\pi)$.

In the string picture, we denote by $a_1$ and $a_2$ the position of each D5 brane along $X_6$, and by $M$ the position of the D3 brane along $X_6$. It follows that the masses of the two fermions are $a_1-M$ and $a_2-M$. , with $a_2 = - a_1$.\\

The instanton moduli space of the theory is a priori singular, but it can be regularized with the introduction of the $\Omega$-background \cite{Nekrasov:2002qd,Nekrasov:2003rj}. Namely, we turn on (real) chemical potentials $\epsilon_1$ and $\epsilon_2$ for the rotation of the $\bR^2_{\text{\tiny{12}}}$ and $\bR^2_{\text{\tiny{34}}}$ planes, respectively. Let us introduce $z_1$ to denote the complex coordinate on $\mathbb{R}^2_{12}$, and $z_2$ to denote the complex coordinate on $\mathbb{R}^2_{34}$. Then, we can view the 5d spacetime as a $\mathbb{R}^2\times \mathbb{R}^2$ bundle over $S^1(R)$, where as we go around the circle,  we make the identification
\begin{align}\label{omega}
(z_1, z_2) \sim  (z_1\, e^{R \epsilon_1}, z_2\, e^{R \epsilon_2})\; .
\end{align}
The resulting space, which we denote as $S^1(R)\times\mathbb{R}^4_{\E_1,\E_2}$, is called the 5d $\Omega$-background.\\

\vspace{10mm}

------\; {\bf Evaluation of the Partition Function}\; ------\\

We are now ready to derive non-perturbative Schwinger-Dyson equations for the $SU(2)_{\pi}$ theory. As we motivated earlier, this amounts to computing the partition function of the theory on $S^1(R)\times\mathbb{R}^4_{\E_1, \E_2}$ in the presence of the loop defect. 

Such a partition function is, by construction, defined for a $U(2)$ gauge theory instead of a $SU(2)$ one, so we will need to decouple the overall $U(1)$ by hand (in the brane picture, this corresponds to the center of mass of the two D5 branes, which is supposed to be a massive mode and decouple). This is an important distinction for us, since we will want to interpret the $SU(2)$ theta angle as a $U(2)$ Chern-Simons level. Then, from now on, let us take the gauge group to be $G=U(2)$, with Chern-Simons level $\kappa\in\{0, 1\}$, and we will specialize to $SU(2)$ with $\kappa=1$ only at the very end\footnote{One could worry about a global anomaly due to the $U(1)\subset U(2)$, but our brane realization of the Wilson loop actually engineers a bare 1d Chern-Simons term at level 1/2, which makes the system anomaly-free. This is essentially because of the flux due to the D3 brane on the D5 branes \cite{Assel:2015oxa,Danielsson:1997wq}.}.
The partition function can be expressed as the Witten index of the $U(k)$  $\cN=(0,4)$ gauged quantum mechanics we just reviewed \cite{Hwang:2014uwa,Kim:2016qqs}. The index can be evaluated using equivariant localization, and will depend on the following fugacities: 

-- The instanton counting parameter, which we will call $\widetilde{\fq}$, associated to the topological $U(1)$ symmetry. It implicitly contains the 5d Yang-Mills coupling.

-- The rotation parameters $e^{\epsilon_1}$ and $e^{\epsilon_2}$ of $\mathbb{R}^4_{\E_1, \E_2}$. They are associated to symmetry generators $J_1+J_R$ and $J_2+J_R$, where $J_1$ and $J_2$ are Cartan generators of $SO(4)_{1234}=SU(2)_1\times SU(2)_2$, and $J_R$ is the Cartan generator of the $R$-symmetry $SO(3)_{789}=SU(2)_R$.  The following notation will come in handy: 
\beq
\label{epsilons}
\epsilon_+\equiv \frac{\epsilon_1+\epsilon_2}{2}\; , ~~~~~ \epsilon_-\equiv \frac{\epsilon_1-\epsilon_2}{2}\; .
\eeq

-- The 5d Coulomb parameters $a_{1, 2}$ of $U(2)$.

-- The defect parameter $M$ associated to $U(1)_{loop}$. Recall that the masses of the two corresponding fermions were identified earlier as $a_1-M$ and $a_2-M$.

-- Finally, the Chern-Simons level $\kappa$ of $U(2)$. The 5d Chern-Simons term  induces a 1d Chern-Simons term on the $U(k)$ quantum mechanics, which is compatible with $(0,4)$ supersymmetry \cite{Collie:2008vc,Kim:2008kn}.

The above symmetry groups are all global symmetries from the point of view of the $U(k)$ quantum mechanics.\\

The index is the grand canonical ensemble of instanton BPS states. It factorizes into the product of a perturbative factor involving the classical and 1-loop contributions, and of a factor capturing the instanton corrections: 
\beq
Z^{full}(M)=Z^{pert}\cdot Z^{inst}(M) \; .
\eeq
Only the factor $Z^{inst}(M)$ depends on the defect fugacity $M$.\\

\begin{figure}[h!]
	%\begin{center}
	\emph{}
	%\hspace{-20ex}
	\centering
	\includegraphics[trim={0 0 0 3cm},clip,width=0.8\textwidth]{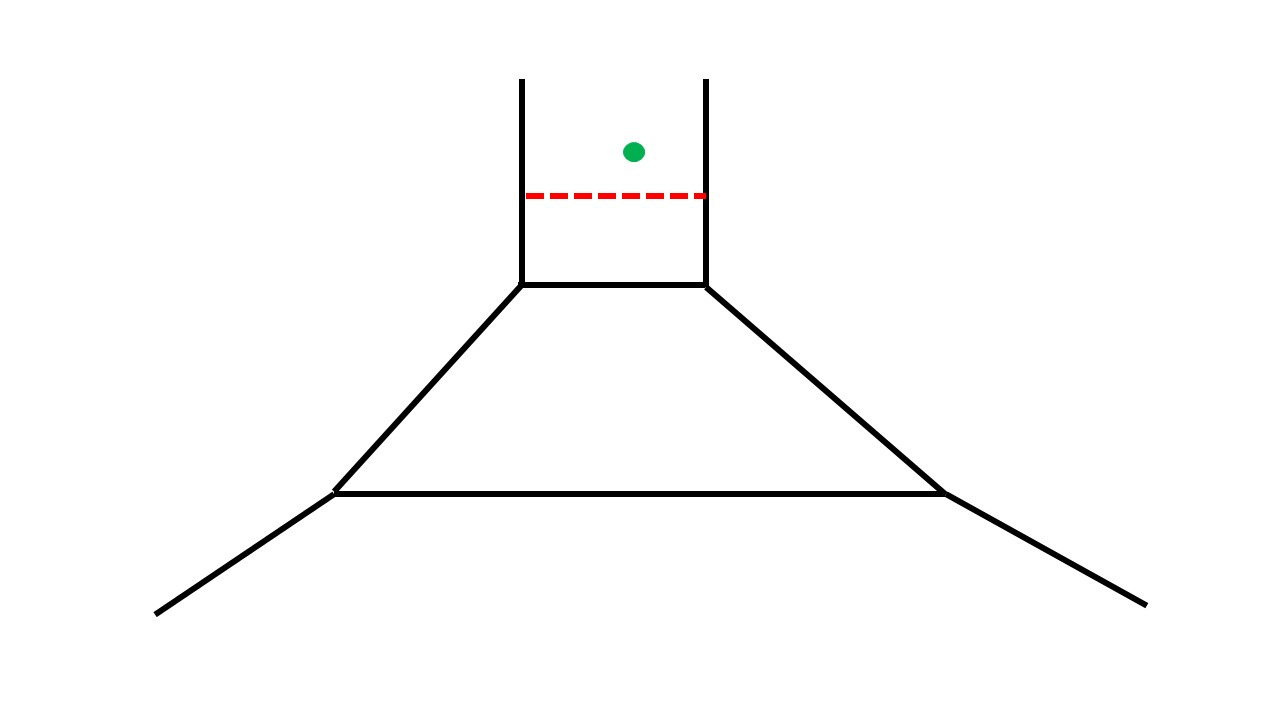}
	\vspace{-12pt}
	\caption{The $U(2)$ gauge theory at Chern-Simons level $\kappa=2$, and a D1  instanton string as a red segment between the two parallel NS5 branes. Because the D3 brane is also placed between the two NS5 branes, this configuration will result in extra UV contributions to the partition functions, $Z^{extra}\neq 1$, which need to be factored out.} 
	\label{fig:E1level2}
	%\end{center}
\end{figure}

A subtle but important question is whether the UV quantity $Z^{inst}(M)$  captures the true instanton counting of the low energy QFT. Indeed, when counting instantons, we are only interested in the Higgs branch of the quantum mechanics. It could happen that the Higgs branch meets the Coulomb branch at a location in moduli space where the Coulomb branch develops a continuum, resulting in extra spurious states. These will manifest themselves in 5d as  decoupled bulk states. We collect these extra unwanted contributions in a factor $Z^{extra}$:
\beq
\label{extra}
Z^{inst}(M) \equiv Z(M)\cdot Z^{extra}(M)
\eeq
An underlying string theory picture such as the one we provided usually helps to identify $Z^{extra}(M)$. For instance, consider our gauge theory $U(2)$ at Chern-Simons level $\kappa=2$. This should be equivalent to the theory at $\kappa=0$, but the index is known to be different \cite{Bergman:2013ala}. The reason is clear in our $(p,q)$ web, since when $\kappa=2$, there are two parallel NS5 branes, which means D1 branes between them can escape to infinity in the $X_6$ direction. This hints at a continuum in the Coulomb branch of the quantum mechanics, see figure \ref{fig:E1level2}. As a result, when the D3 brane happens to be located between the NS5 branes, there will be unwanted D3/D1 string states that are counted in the index. In this section, our goal is to eventually set $\kappa=1$, and in that case, it will turn out that $Z^{extra}=1$. We will later see that in the $Sp(1)$ formalism, $Z^{extra}\neq 1$.\\

In order to make contact with Seiberg-Witten geometry, we actually need to normalize the index by the partition function of the pure $U(2)$ theory, in the absence of loop defect. We denote it as
\beq
Z^{full}_{\text{pure}}=Z^{pert}\cdot Z^{inst}_{\text{pure}} \; ,
\eeq
where the instanton part once again contains a QFT contribution and potentially an extra contribution due to the UV completion:
\beq
\label{extrapure}
Z^{inst}_{\text{pure}}\equiv  \langle 1 \rangle \cdot Z^{extra}_{\text{pure}} \; .
\eeq
Notice that $Z^{pert}$ cancels out after normalization, so we will not need to compute it:
\beq
\label{extranormalized}
\frac{Z^{full}(M)}{Z^{full}_{\text{pure}}} = \frac{Z^{inst}(M)}{Z^{inst}_{\text{pure}}} = \frac{Z(M)}{\langle 1 \rangle} \cdot \frac{Z^{extra}(M)}{Z^{extra}_{\text{pure}}} \, .
\eeq
The ratio $\frac{Z(M)}{\langle 1 \rangle}$ turns out to have a remarkable regularity property in the variable $e^M$.
To understand this, we organize the instanton contribution $Z^{inst}(M)$ as a sum over all instanton sectors $k$, where the integer $k=\frac{1}{8\pi^2}\int_{\mathbb{R}^4}\mbox{Tr}(F\wedge F)$ is the $U(1)$ topological charge in 5d, conjugate to the instanton counting fugacity $\widetilde{\fq}$.\\

We can evaluate the gauge theory index in the weak coupling regime of the  UV quantum mechanics, where it reduces to Gaussian integrals around saddle points.  These saddle points are parameterized by  $\phi=R \varphi^{(QM)}+ i R A^{(QM)}_t$, with $A^{(QM)}_t$ the gauge field and $\varphi^{QM}$ the scalar in the vector multiplet of the quantum mechanics, and $R$ the radius of the $S^1$ along $X_0$. Since the gauge group of the quantum mechanics is $U(k)$, we denote the (complexified) eigenvalues of $\phi$ as $\phi_1, \ldots, \phi_k$. Performing the Gaussian integrals over massive fluctuations, the index reduces to a zero mode integral of various 1-loop determinants: 
\begin{align}
\label{5dintegralA1}
& Z^{inst}(M)  =\sum_{k=0}^{\infty}\;\frac{\widetilde{\fq}^{k}}{k!} \, \oint \left[\frac{d\phi_I}{2\pi i}\right]Z^{(k)}_{vec}\cdot Z^{(k)}_{C.S.}\cdot  Z^{(k)}_{defect}(M)  \; , \\
&Z^{(k)}_{vec} =\prod_{\substack{I\neq J\\ I,J=1}}^k\sh(\phi_I-\phi_J)\prod_{I, J=1}^k\frac{\sh(\phi_I-\phi_J+2\E_+)}{\sh(\phi_I-\phi_J+\E_+\pm\E_-)}\prod_{I=1}^k\prod_{i=1}^2\frac{1}{\sh(\pm(\phi_I-a_i)+\E_+)}\; ,\nonumber\\
&Z^{(k)}_{C.S.} =(-1)^{k\, \kappa}\prod_{I=1}^k e^{\phi_I\, \kappa}\; ,\nonumber\\
&Z^{(k)}_{defect} =\prod_{i=1}^2\sh(a_i-M) \prod_{I=1}^k\frac{\sh(\pm(\phi_I-M)+\E_-)}{\sh(\pm(\phi_I-M)-\E_+)}\, .\nonumber
\end{align}
Above, we defined a convenient notation $\sh(x)\equiv 2\sinh(x/2)$, where products over all signs inside an argument should be considered; for example, $\sh(a\pm b)\equiv\sh(a+b)\,\sh(a-b)$. The presence of $k!$ in the denominator is the Weyl group order of $U(k)$.\\ 

-- The factor $Z^{(k)}_{vec}$ contains the physics of the 5d vector multiplet.
 
-- The factor $Z^{(k)}_{defect}$ contains the physics of the Wilson loop defect.

-- The factor $Z^{(k)}_{C.S.}$ contains the physics of the $U(2)$ Chern-Simons term
\beq
\frac{\kappa}{24\, \pi^2}\int Tr\left[A\wedge F \wedge F + \frac{i}{2} A^3\wedge F -\frac{1}{10} A^5 \right] \; ,
\eeq
with $\kappa$ an integer\footnote{A more general statement is that if the 5d theory has an even number of fundamental hypermultiplets, then $\kappa$ is integer, while if the theory has an odd number of fundamental hypermultiplets, $\kappa$ is half-integer.}. Note the presence of the phase $(-1)^{k\, \kappa}$, which has been argued  to be due to the decoupling of the overall $U(1)$ inside $U(2)$ \cite{Bergman:2013ala}. Indeed, viewing $U(2)$ as $U(1)\times SU(2)$, the bare $U(2)$ Chern-Simons term becomes a mixed Chern-Simons term for the $SU(2)$ theory, proportional to
\beq
\kappa\int A\wedge Tr\left[F\wedge F\right] 
\eeq
This explains the dependence on $k$ in  $(-1)^{k\, \kappa}$.\\

The partition function is a multi-dimensional integral over the 1d Coulomb moduli $\phi_I$'s, which can be evaluated by residues using an ``$i \E$"  prescription, or using the so-called Jeffrey-Kirwan residue \cite{Jeffrey:1993}. For a review of the latter in our context, see \cite{Hwang:2014uwa,Cordova:2014oxa,Hori:2014tda} and \cite{Benini:2013xpa}.\\

The loop defect factor $Z^{(k)}_{defect}$  is made up of two contributions, the first of which is
\beq
\prod_{i=1}^2\sh(a_i-M)\; .
\eeq
These are the 1-loop determinants obtained from the quantization of the D3/D5 strings in the web. Such a quantization was first carried out in our context in \cite{Gomis:2006sb}, see also \cite{Banks:1997zs}. This is the contribution of 2 Fermi multiplets with a single fermion, transforming in the bifundamental representation of $U(2)\times U(1)_{loop}$. Second, there are factors
\beq
\prod_{I=1}^k\frac{\sh(\pm(\phi_I-M)+\E_-)}{\sh(\pm(\phi_I-M)-\E_+)} \; ,
\eeq
which are the 1-loop determinants obtained from the quantization of the D3/D1 strings. It is the contribution of one fundamental twisted hypermultiplet and one Fermi multiplet with two fermions.

Note that there are more string configurations involving the D3 brane that we have described so far. For instance,  one could have D1 strings stretching between the D3 brane and the NS5. However, such a sector has to do with 't Hooft loop physics instead \cite{Brennan:2018rcn,Brennan:2018moe}, so we should decouple it to start with. All in all, $Z^{(k)}_{defect}$ above contains exclusively Wilson loop physics.\\

Now, the key step is to define a defect loop operator vev,
\begin{align}
\label{YoperatorA1}
\left\langle \left[Y(M)\right]^{\pm 1}\right\rangle \equiv
\sum_{k=0}^{\infty}\frac{\widetilde{\fq}^{k}}{k!} \, \oint_{pure}  \left[\frac{d\phi_I}{2\pi i}\right]Z^{(k)}_{vec}\cdot Z^{(k)}_{C.S.}\cdot \left[Z^{(k)}_{defect}(M)\right]^{\pm 1}\, . 
\end{align}
At first sight, the one-point function $\left\langle Y(M)\right\rangle$ looks like the partition function, but they differ in the choice of contours. The contour integral for the $Y$-operator vev is \emph{defined} to avoid all poles from the defect factor $Z^{(k)}_{defect}$. In other terms, since the dependence of the integrand on the fugacity $M$ is exclusively due to $Z^{(k)}_{defect}$, none of the poles will depend on $M$.
 
This is very different from the partition function, where either the ``$i \E$" prescription or the Jeffrey-Kirwan residue order us to enclose exactly one pole coming from the defect factor $Z^{(k)}_{defect}$:
\begin{equation}
\label{onlyonepole}
\phi_I-M-\epsilon_+=0 \;\;\mbox{for some} \; \text{for some}\; I\in\{1,\ldots, k\} \; .
\end{equation}

Then, we find by direct computation that the partition function can be written in terms of the $Y$-operator loop vevs, as a sum of exactly two terms:
\begin{align}
\label{A1pure}
Z^{inst}(M) = \left\langle Y(M) \right\rangle + \widetilde{\fq} \; (-1)^\kappa\, e^{\kappa(M+\E_+)} \left\langle\frac{1}{Y(M+2\,\epsilon_+)}\right\rangle\; .
\end{align}
where $\kappa\in\{0,1\}$.
The meaning of this expression is as follows: the first term on the right-hand side encloses almost all the (physical) poles of the partition function integrand, but there is exactly one pole missing: the extra pole at $\phi_I=M+\epsilon_+$ from $Z^{(k)}_{defect}$. Nontrivially, the residue at this missing pole is equal to the vev of the $Y$-operator inverse, with a shift: $(-1)^\kappa\, e^{\kappa(M+\E_+)}\; \left\langle Y(M + 2\, \epsilon_+)^{-1}\right\rangle$. This is proved by simply evaluating the index integral \eqref{5dintegralA1pure}, and using the integral definition of the $Y$-operator \eqref{YoperatorA1}. The presence of the parameter $\widetilde{\fq}$ in the second term counts exactly one instanton, to make up for the missing $M$-pole.\\

We now decouple the overall $U(1)$ by setting $a_1 = - a_2 \equiv a$, and further set $\kappa=1$. We obtain at once the partition function of the $SU(2)_{\pi}$ theory in the presence of a fundamental Wilson loop:
\beq
Z^{inst}_{\pi}(M) = \left\langle Y(M) \right\rangle - \widetilde{\fq} \;  e^{M+\E_+} \left\langle\frac{1}{Y(M+2\,\epsilon_+)}\right\rangle \; .
\eeq

To make contact with Seiberg-Witten geometry, we normalize the partition function by the pure contribution,
\begin{align}
\label{5dintegralA1pure}
& Z^{inst}_{\text{pure}, \pi} =\sum_{k=0}^{\infty}\;\frac{\widetilde{\fq}^{k}}{k!} \, \oint_{pure}  \left[\frac{d\phi_I}{2\pi i}\right]Z^{(k)}_{vec}\cdot Z^{(k)}_{C.S.} \; .
\end{align}
Because there are no spurious UV contributions coming from our ADHM realization, we rewrite the normalized index $Z^{inst}_{\pi}(M)/Z^{inst}_{\text{pure}, \pi}$  as $Z_{\pi}(M)/\left\langle 1 \right\rangle$, following our previous notation \eqref{extranormalized}. Expanding this partition function in the (exponentiated) defect fugacity $e^{M}$, we find
\beq
\label{SchwingerA1pure}
\boxed{\frac{1}{\left\langle 1 \right\rangle}\left[	\left\langle Y(M) \right\rangle - \widetilde{\fq} \; e^{(M+\E_+)} \left\langle\frac{1}{Y(M+2\,\epsilon_+)}\right\rangle\right]  = e^{-M}\,\cT(M; \{\beta_i^{\epsilon_1, \epsilon_2}\})} \; ,
\eeq
where $\cT(M; \{\beta_i^{\epsilon_1, \epsilon_2}\})$ is a \emph{finite} polynomial in $e^M$, whose roots $\{\beta_i^{\epsilon_1, \epsilon_2}\}$ define the vacuum.  To prove this is the case, it suffices to show that $\cT(M; \{\beta_i^{\epsilon_1, \epsilon_2}\})$ is regular in $e^M$. This is most easily done by looking at the (lack of) $M$-related poles in the integrand which can pinch a given contour. From the explicit expressions \eqref{5dintegralA1}, this is a quick exercise\footnote{Details can be found for instance in \cite{Kim:2016qqs} for $U(N)$ at $\kappa=0$. The same reasoning applies here, since the presence of the Chern-Simons term does not introduce any $M$-poles inside the integrand.}. Furthermore, the asymptotics of $Z^{inst}(M)$ at $M\rightarrow\pm\infty$  indicate the polynomial is monic and of order 2.

The regularity of the partition function in $e^M$ is a proof that the boxed expression is a non-perturbative Schwinger-Dyson identity for the $SU(2)_{\pi}$ gauge theory, solved explicitly by the $Y$-operator vev \eqref{YoperatorA1}.\\

At this point, we find it convenient to switch to exponentiated variables, so we define\footnote{The radius $R$ of the $X_0$-circle is implicit in all the exponents, and we have set it to 1 in all expressions. It is an easy task to reinstate it explicitly if needed.}:
\begin{align}
\label{5dvariables}
&z\equiv e^M \;\; , \;\;\;\;\; \alpha\equiv e^a \;\; , \;\;\;\;\; q\equiv e^{\E_1} \;\; , \;\;\;\;\; t\equiv e^{-\E_2}\\
&v\equiv\sqrt{\frac{q}{t}}=e^{\E_+} \;\; , \;\;\;\;\; u\equiv \sqrt{q\, t}=e^{\E_-}\nonumber
\end{align}
The Schwinger-Dyson identity takes the form
\beq
\label{Schwinger5d}
\frac{1}{\left\langle 1 \right\rangle}\left[\left\langle Y(z) \right\rangle - \widetilde{\fq} \; z \, v\, \left\langle\frac{1}{Y(z\,v^2)}\right\rangle\right]  = z^{-1}\, \cT(M; \{\beta_i^{\epsilon_1, \epsilon_2}\})\; .
\eeq
Explicitly, the right-hand side has the following form
\beq
\label{RHSschwinger}
z^{-1}\,\cT(z; \{\beta_i^{q, t}\}) = z - \left(   \alpha+\alpha^{-1}+\ldots   \right)   + z^{-1} 
\eeq
The dotted terms stand for higher instanton corrections, which do not depend on the defect fugacity $z$. The above identity has several important implications, which we now turn to.

\vspace{10mm}

------\; {\bf Physics of the Schwinger-Dyson Identity}\; ------\\

First, it has been conjectured that non-perturbative Schwinger-Dyson equations for all classical gauge groups can be understood as a double quantization of Seiberg-Witten geometry  \cite{Haouzi:2020yxy}. A necessary condition for this to be true is that in the flat space limit $q, t \rightarrow 1$, one should recover the Seiberg-Witten curve of the theory. In the case at hand, we must first reinterpret the $Y$-operators in flat space as complex coordinates on the cylinder. Namely, define the map \cite{Nekrasov:2012xe}
\begin{align}
\label{ycoordinates}
y \; : \; &\mathbb{C}^\ast\rightarrow \;\mathbb{C}^\ast\\
& z \;\;\mapsto \;\left\langle Y(z) \right\rangle
\end{align}
We see at once that after turning off the $\Omega$-background, the Schwinger-Dyson identity becomes the Seiberg-Witten curve of  5d $SU(2)_{\pi}$, with coordinates $(z,y)\in\mathbb{C}^\ast\times\mathbb{C}^\ast$\footnote{The Seiberg-Witten curve can be for instance derived directly from our $SU(2)_{\pi}$ brane web, see the appendix of \cite{Kim:2014nqa}.}:
\beq
\label{SWcurve}
y(z) - \widetilde{\fq}\, \frac{z}{y(z)} = z^{-1}\,\cT(z; \{\beta_i^{1, 1}\})
\eeq
Above, $\cT(z; \{\beta_i^{1, 1}\})$, is still a degree 2 monic polynomial in $z$; we simply took the flat space limit $q, t \rightarrow 1$ of the roots $\{\beta_i^{q, t}\}$.\\

Second, the left-hand side of the identity has an interpretation in representation theory, as a deformed character of the spin 1/2 representation of the quantum affine algebra $U_q(\widehat{A_1})$. Following \cite{Nekrasov:2015wsu}, we call it a fundamental $qq$-character. It reads
\beq
\label{qqcharacter}
\chi^{(A_1, \pi)}_{qq}=\frac{1}{\left\langle 1 \right\rangle}\left[\left\langle Y(z) \right\rangle - \widetilde{\fq} \; z \, v\, \left\langle\frac{1}{Y(z\,v^2)}\right\rangle\right] \; ,
\eeq
where the second term is built from the first through the ``i-Weyl" \cite{Nekrasov:2012xe} group action
\beq
\label{iWeyl}
\left\langle Y(z) \right\rangle  \mapsto  - \widetilde{\fq} \; z \, v\, \left\langle\frac{1}{Y(z\,v^2)}\right\rangle \; .
\eeq
The regularity of this character is guaranteed by the Schwinger-Dyson identity, namely the finiteness of the Laurent polynomial \eqref{RHSschwinger}. Note the additional twist in the definition of the i-Weyl action compared to the usual $\theta=0$ case: in that case, the $qq$-character
\beq
\label{qqcharacterzero}
\chi^{(A_1, 0)}_{qq}=\frac{1}{\left\langle 1 \right\rangle}\left[\left\langle Y(z) \right\rangle + \widetilde{\fq} \; \left\langle\frac{1}{Y(z\,v^2)}\right\rangle\right] 
\eeq
is generated through the i-Weyl group action
\beq
\label{iWeylzero}
\left\langle Y(z) \right\rangle  \mapsto   \widetilde{\fq} \;  \left\langle\frac{1}{Y(z\,v^2)}\right\rangle \; .
\eeq
\\

\begin{figure}[h!]
	%\begin{center}
	\emph{}
	%\hspace{-20ex}
	\centering
	\includegraphics[trim={0 0 0 0.5cm},clip,width=0.8\textwidth]{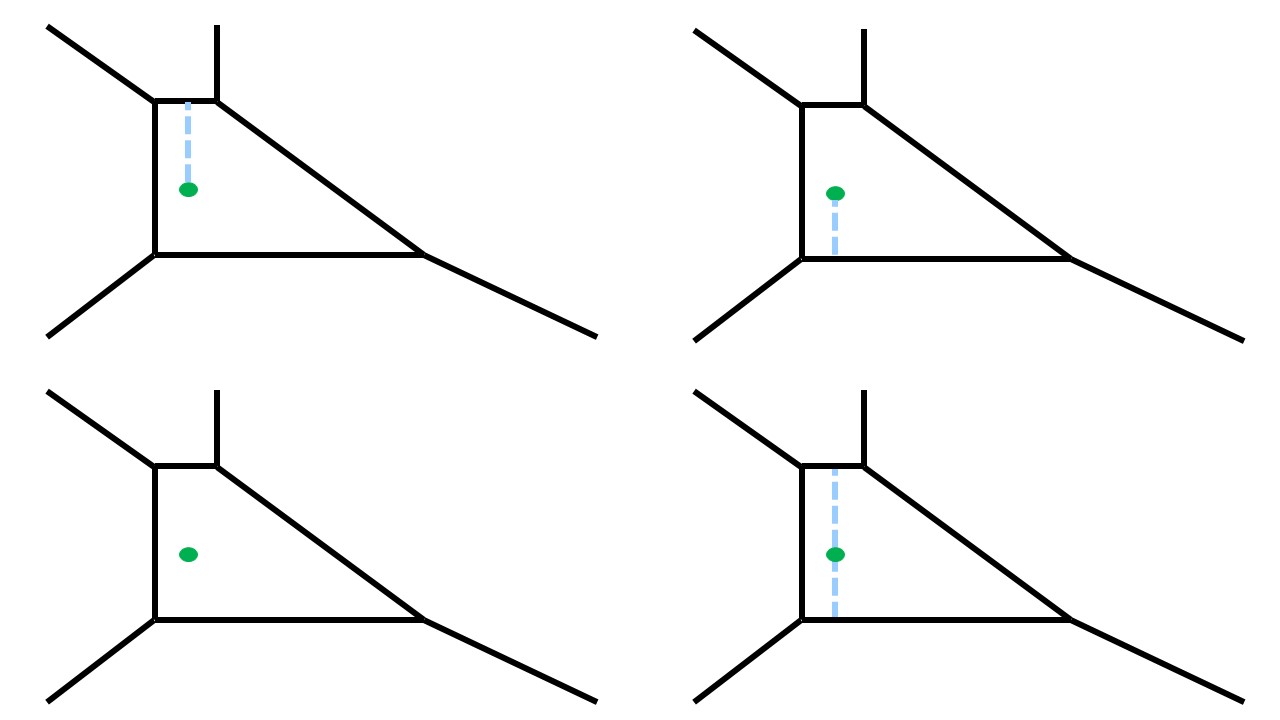}
	\vspace{-4pt}
	\caption{The partition function as a generating function of Wilson loop vevs, from the $(p,q)$ web perspective. Since there can be at most one fundamental string stretched between a D5 and a D3 brane (the zero mode being fermionic), we are effectively summing over the four configurations above. The top two configurations carry an almost trivial contribution to the partition function, as can be seen for example using a Hanany-Witten move \cite{Hanany:1996ie}. In the top left picture, we can pull the D3 brane up and past the top D5 brane, resulting in a decoupled D3 brane. The flux on the D5 branes due to the D3 brane $U(1)_{loop}$ results in a classical contribution $z$ to the partition function (instead of 1). Similarly for the top right picture, pulling the D3 brane down and past the bottom D5 brane, the D3 brane decouples from the web, and the only contribution there is again due to the induced flux, which contributes the term $z^{-1}$. The bottom two pictures are configurations with no net fundamental strings ending on the D3 brane, which means the corresponding  states are neutral under $U(1)_{loop}$. The bottom left picture represents a classical contribution  $\alpha$ to the partition function, while the bottom right represents a classical contribution  $\alpha^{-1}$. Note that in those two cases, the D3 brane cannot be decoupled from the web, so we expect nontrivial instanton corrections due to the $k$ D1 strings. On the other hand, because the D3 brane is decoupled in the top two pictures, the classical contributions $z$ and $z^{-1}$ are expected to be exact.} 
	\label{fig:E1tilde4cases}
	%\end{center}
\end{figure}

Third, the right-hand side of the Schwinger-Dyson identity encodes Wilson loop vev physics. Indeed, it is known that the partition function of the gauge theory on D5 branes in the presence of a D3 brane loop defect is the generating function of Wilson loop vevs, in all fundamental representations of the gauge group \cite{Gomis:2006sb}. In our case, the gauge group is $SU(2)_{\pi}$, so there is a unique fundamental representation, the spin 1/2 representation. We expect from the brane web that the Wilson loop vev should be the $U(1)_{loop}$-neutral term in the partition function, see figure \ref{fig:E1tilde4cases}. Put differently, the Wilson loop vev is the $z$-independent term in the right-hand side of our Schwinger-Dyson identity,
\beq
\label{RHSwilson}
z^{-1}\,\cT(z; \{\beta_i^{q, t}\})= z - \langle W_{\bold{2}} \rangle + z^{-1} \; .
\eeq

In the absence of D1 strings, the above is nothing but the 1-loop determinant obtained from quantized D5/D3 strings, and the Wilson loop vev is simply $\langle W_{\bold{2}} \rangle=\alpha+\alpha^{-1}$. In our ADHM quantum mechanics, the vev gets an infinite number of instanton corrections. Explicitly, we expand the left-hand side of \eqref{Schwinger5d} in $z$ and identify the $U(1)_{loop}$-neutral term to be:
\beq
\label{Wilsonexplicit}
\langle W_{\bold{2}} \rangle = \alpha+\alpha^{-1} +\widetilde{\fq}\,\frac{\alpha^2\,(v+v^{-1})}{(1-(v\,\alpha)^2)(1-(\alpha/v)^2)} +o\left[\widetilde{\fq}^2\right]
\eeq
The expansion can be carried out to an arbitrary high instanton order, by performing an increasing number of residue integrals. It is illuminating to introduce new fugacities\footnote{The notation comes from Calabi-Yau geometry, where we can assign K\"{a}hler classes to the fiber and base $\mathbb{P}^1$ for the Hirzebruch surface $\mathbb{F}_1$; this is the geometry relevant to the $SU(2)_{\pi}$ theory.}:
\begin{align}
\label{5dvariablesWilson}
&\alpha\equiv Q^{1/2}_F \;\; , \;\;\;\;\; \widetilde{\fq}\equiv - Q_B \, Q_F^{-1/2}
\end{align}
Then, we can present the Wilson loop vev as a BPS-like expansion, up to normalization:
\begin{align}
\label{WilsonBPS}
Q^{1/2}_F\, \langle W_{\bold{2}} \rangle =  1 &+ Q_F + Q_B\, Q_F\, \chi^{SU(2)}_{\bold{2}}(v) +  Q_B\, Q^{2}_F \, \chi^{SU(2)}_{\bold{4}}(v)\\
& +  Q_B\, Q^{3}_F \, \chi^{SU(2)}_{\bold{6}}(v) +  Q_B\, Q^{4}_F \, \chi^{SU(2)}_{\bold{8}}(v) +  Q^2_B\, Q^{2}_F\, \chi^{SU(2)}_{\bold{5}}(v)\nonumber\\ 
& +   Q^2_B\, Q^{3}_F\, \left(\chi^{SU(2)}_{\bold{2}}(u)\, \chi^{SU(2)}_{\bold{8}}(v)+2\, \chi^{SU(2)}_{\bold{7}}(v)+\chi^{SU(2)}_{\bold{5}}(v)\right)\nonumber\\
& + \ldots \nonumber
\end{align}
Above, we wrote $\chi^{SU(2)}_{\bold{j}}$ for the character of the j-dimensional irreducible representation of $SU(2)$; for instance, the characters $\chi_{\bold{2}}(u)=\sqrt{q\,  t} + \frac{1}{\sqrt{q \, t}}$ and $\chi_{\bold{5}}(v)=\frac{q^2}{t^2} +\frac{q}{t} + 1 + \frac{t}{q} + \frac{t^2}{q^2}$ appear in the expansion.
A geometric interpretation of this expansion is currently lacking; it would be important to understand it properly.\\

A similar analysis of the Wilson loop vev when $\theta=0$ was carried out in detail in \cite{Assel:2018rcw}. The Wilson loop vev in that case comes out to be
\beq
\label{Wilsonexplicitzero}
\langle W_{\bold{2}} \rangle = \alpha+\alpha^{-1} -\widetilde{\fq}\,\frac{v^2\,(\alpha+\alpha^{-1})}{(1-(v\,\alpha)^2)(1-(v/\alpha)^2)} +o\left[\widetilde{\fq}^2\right] \; ,
\eeq
as can be explicitly checked by taking $\kappa=0$ in \eqref{A1pure}.
The expansion of the $SU(2)_0$ loop vev in clever (meaning S-duality invariant)  coordinates hints at an enhancement of the topological $U(1)$ global symmetry to $E_1 = SU(2)$ at the fixed point. From our expression \eqref{WilsonBPS}, it is clear that no such enhancement occurs in our $SU(2)_{\pi}$ analysis, which is consistent with the fact that the fixed point theory  $\widetilde{E_1}$ only has a $U(1)$ global symmetry.\\

We now give a second presentation of the $SU(2)_{\pi}$ quantum geometry, this time from an ADHM quantum mechanics originating in type I' string theory.

\section{The $Sp(1)_{\pi}$ quantized geometry, the type I' way}

Consider type IIA string theory in flat space. We compactify the $X_0$-direction on a circle $S^1(R)$ of radius $R$, introduce an orientifold O$8^-$ plane and  the following branes:
\begin{table}[H]\centering
	\begin{tabular}{|l|cccccccccc|} \hline
		&  0 &  1 &  2 &  3 &  4 &  5 &  6 &  7 &  8 &  9\\
		\hline\hline
		\;\;O$8^-$ &\xTB& \xTB  &\xTB   & \xTB  & \xTB  &  \xTB  &  \xTB  & \xTB   & \xTB   &    \\
		\hline
		$k$\,\,	D0 &\xTB&   &   &   &   &    &    &    &    &    \\
		\hline
		$1$\,\,	D4 &\xTB&\xTB&\xTB&\xTB&\xTB&    &    &    &    &     \\
		\hline
		$1$\,\, D$4'$ &\xTB&    &    &    &    &\xTB&\xTB&\xTB&\xTB&  \\
		\hline
		\;	F1 &\xTB&    &    &    &    & & & & & \xTB   \\
		\hline
	\end{tabular}
	\caption{The directions of the various branes. The direction $X_0$ is compact.}
	\label{table:BranesSpN}
\end{table}

Due to the presence of the orientifold plane, the low energy effective theory on the D4 brane is a 5d $Sp(1)$ gauge theory on $S^1(R)\times\mathbb{R}^4$. On the Coulomb branch, the distance between a D4 brane and its image (the Coulomb modulus) is defined to be $2 a$.

The D$4'$ brane wraps the $X_0$-circle and  sits at the origin of $\mathbb{R}^4$. As such, it realizes a 1/2-BPS loop defect from the point of view of the D4 brane theory. Finally, we think of the $k$ D0 branes as instantons of the D4 brane theory. The gauge group on the D0 branes is $\widehat{G}=O(k)$, which has a $\mathbb{Z}_2$ center. As a set, it is $O(k)^+\cup O(k)^-$, where the superscript is the charge under this $\mathbb{Z}_2$. The component $O(k)^+$ is the group $SO(k)$, while $O(k)^-$, the set of elements of $O(k)$ with determinant -1, does not form a group.\\

In gauge theory terms, the D$4'$ brane theory is coupled to the D4 brane theory once again through 1d fermions along $X_0$, with associated ``defect" group $Sp(1)_{loop}$. Therefore, the fermions transform in the bifundamental representation of $Sp(1)\times Sp(1)_{loop}$, where the first group is the D4 brane group, and the second group is the D$4'$ brane group. Pulling the  D$4'$ brane a distance $M$ away in the $X_9$-direction, there are now open strings with nonzero tension between the D4 and D$4'$ brane; the resulting fermions can then be considered heavy probes with mass proportional to the distance $M$. The action term coupling the 1d fermions to the bulk is as before,
	\begin{equation}
\label{5d1ddfermionSp1}
S^{5d/1d}=\int dt\; \chi_{i,\rho}^\dagger\, \left( \delta_{\rho\sigma}(\delta_{ij} \, \partial_t - i\,A^{[5d]}_{t, ij}   + \Phi^{[5d]}_{ij})   - \delta_{ij}\, \widetilde{A}_{t,\rho \sigma} \right)\, \chi_{j,\sigma} \; .
\end{equation}
Above, $A^{[5d]}_t$ and $\Phi^{[5d]}$ are the pullback of the 5d gauge field and the adjoint scalar of the vector multiplet, respectively.  $\widetilde{A}_{t}$ is the (nondynamical) gauge field the 1d fermions couple to, with eigenvalues $\pm M$. The indices $i$ and $j$ are  in the fundamental representation of $Sp(1)$, while the indices $\rho$ and $\sigma$ are  in the fundamental representation of $Sp(1)_{loop}$. The variable $t$ is periodic, with period $R/(2\pi)$.\\

Finally, we put the theory on the 5d $\Omega$-background $S^1(R)\times\mathbb{R}^4_{\E_1,\E_2}$. The following fugacities which will come in handy:
\ie
\label{Sp1fugacities}
\epsilon_+\equiv{\epsilon_1 + \epsilon_2\over 2},~~~ \epsilon_-\equiv{\epsilon_1 - \epsilon_2\over 2},
~~~ m \equiv {\epsilon_3 - \epsilon_4\over 2},
\fe
where $\epsilon_1$, $\epsilon_2$, $\epsilon_3$, and $\epsilon_4$ are the chemical potentials associated to the rotations of the $\bR^2_{\text{\tiny{12}}}$, $\bR^2_{\text{\tiny{34}}}$, $\bR^2_{\text{\tiny{56}}}$, and $\bR^2_{\text{\tiny{78}}}$ planes, respectively.\\

Our interest lies in computing the index of the $O(k)$ ADHM gauged quantum mechanics on the D0 branes, which in our background has $(0,4)$ supersymmetry. The explicit field content of of the quantum mechanics is given for instance in \cite{Chang:2016iji,Haouzi:2020yxy}. For our purposes, we will only need the various 1-loop determinants that make up the index, and we collected those in the appendix.\\

An important point is that in the absence of D8 branes, the charge of the O8 plane is not canceled, which means the dilaton is running in $X_9$. As a consequence, the $U(1)$ instanton charge $k$ will receive an anomalous contribution, which in turn causes a fractional shift to the instanton charge. The index of the D0 brane quantum mechanics we study here will not be affected by this shift, but other physical quantities typically are (for example, the superconformal index is, as was shown explicitly in \cite{Chang:2016iji}).

\vspace{10mm}

------\; {\bf Evaluation of the Partition Function}\; ------\\

In the previous section, we interpreted the $\theta$-angle of $SU(2)$ as the Chern-Simons level of $U(2)$. What is the origin of the $\theta$-angle for $Sp(N)$? It turns out we can make a discrete choice in the type I' background; consider the RR one-form gauge field $C_1$ in type IIA. After compactifying on the circle along $X_0$, we obtain a new $U(1)$ modulus in 9 dimensions, $\int_{S^1}C_1 = \theta$. The orientifold projection of the O8 plane projects out $\theta$, but not entirely: due to the periodicity $\theta \rightarrow \theta + 2\pi$, the $U(1)$ gets broken to $\mathbb{Z}_2$, so $\theta = 0$ or $\theta=\pi$. From the point of view of the effective theory on the D4 branes, $\theta$ can be regarded as a discrete gauge instanton in 5d, consistent with the mathematical fact $\pi_4(Sp(N))=\mathbb{Z}_2$. From the point of view of the $k$ D0 brane quantum mechanics, $\theta$  can be regarded as a discrete instanton in 1d, consistent with the mathematical fact $\pi_0(O(k))=\mathbb{Z}_2$. There, $e^{i \theta}$ is interpreted a discrete holonomy for the 1d $O(k)$ gauge theory, and correspondingly we have two components $O(k)^{+}$ and $O(k)^{-}$ \cite{Kim:2012gu,Bergman:2013ala}.\\

Once again, we can evaluate the gauge theory index in the weak coupling regime of the  UV quantum mechanics, where it reduces to Gaussian integrals around saddle points. Performing those over massive fluctuations, the index reduces to a zero mode integral of various 1-loop determinants, over the (complexified) Coulomb moduli of the 1d quantum mechanics:
\begin{align}
\label{5dintegralCN}
& Z^{(k,\pm)}(M)  = \frac{1}{|O(k)^{\pm}|^\chi} \, \oint \left[\frac{d\phi_I}{2\pi i}\right]Z^{(k,\pm)}_{vec}\cdot Z^{(k,\pm)}_{antisym}\cdot  Z^{(k,\pm)}_{defect}(M) \; .
\end{align}
We denoted the 1-loop determinants related to $\widehat{G}=O(k)^+$ as ${Z}^{(k,+)}$, and those related to $\widehat{G}=O(k)^-$ as ${Z}^{(k,-)}$. 
For the sake of brevity, we only made the dependence on the defect fugacity $M$ explicit in writing the partition function.
$|O(k)^+|$ is the order of the Weyl group of $SO(k)$, and $|O(k)^-|$ is the order of the Weyl group of $O(k)^-$. We also introduced a notation
\beq
k\equiv 2n+\chi \;\; ,\;\;\;\; \chi=0, 1 \; .
\eeq 	
Explicit expressions for the 1-loop determinants are written in the appendix.\\

-- The factor $Z^{(k,\pm)}_{vec}$ contains the physics of the 5d vector multiplet.

-- The factor $Z^{(k,\pm)}_{antisym}$ contains the physics of a 5d hypermultiplet in the antisymmetric representation of the gauge group. There is an associated mass $m$, which we interpreted geometrically above \eqref{Sp1fugacities}. In what follows, we first evaluate the integrals, and then decouple the antisymmetric matter by taking the limit  $m\rightarrow \infty$.

-- The factor $Z^{(k,\pm)}_{defect}$ contains the physics of the Wilson loop defect. As we noted, it is the only contribution depending on the fugacity $M$.\\

The component $Z^{(k,-)}$ can be understood as inserting a parity operator $(-1)^P$ in the Witten index, with $P$ being the element $-1$ of $O(k)^-$. Note that the 5d Coulomb fugacities are parity-odd, hence the presence of hyperbolic cosines in the 1-loop determinants. To compute a physical partition function, it is necessary to have a spectrum of definite parity. There are two options here, which is precisely the $\mathbb{Z}_2$ choice of $\theta$-angle.

On the one hand, one can project to parity-even states with $\frac{1}{2}(1+(-1)^P)$, which amounts to computing the index 
\beq
\label{Sp1pieces}
Z^{(k)}_0(M) \equiv \frac{1}{2}\left[Z^{(k,+)}(M) + Z^{(k,-)}(M)\right]\;\; , \;\;\;\;\; Z^{inst}_0(M)=\sum_{k=0}^\infty \widetilde{\fq}^{k} \, Z^{(k)}_0(M) .
\eeq
This is the choice $\theta=0$.

On the other hand, one can project to parity-odd states with $\frac{1}{2}(1-(-1)^P)$, which amounts to computing the index 
\beq
\label{Sp1partition}
Z^{(k)}_{\pi}(M) \equiv \frac{(-1)^k}{2}\left[Z^{(k,+)}(M) - Z^{(k,-)}(M)\right]\;\; , \;\;\;\;\; Z^{inst}_{\pi}(M)=\sum_{k=0}^\infty \widetilde{\fq}^{k}\, Z^{(k)}_{\pi}(M) .
\eeq
This is the choice $\theta=\pi$, which is the one of interest for us. There is a priori an ambiguity in the overall sign, but one can argue (indirectly)  that the correct sign is  $(-1)^k$ \cite{Bergman:2013ala}. For instance, when $k=1$, one can show that the ground state of an instanton particle in $Sp(N)_{\pi}$ is fermionic, hence an overall negative sign must be present.\\

To derive non-perturbative Schwinger-Dyson equations, we proceed as in the $SU(2)$ formalism and define a defect loop operator vev, from the building blocks
\begin{align}
\label{5dintegralCNYop}
& \left\langle \left[Y^{(k,\pm)}(M)\right]^{\pm 1} \right\rangle = \frac{1}{|O(k)^{\pm}|^\chi} \, \oint_{pure} \left[\frac{d\phi_I}{2\pi i}\right]Z^{(k,\pm)}_{vec}\cdot Z^{(k,\pm)}_{antisym}\cdot  \left[Z^{(k,\pm)}_{defect}(M)\right]^{\pm 1} \; .
\end{align}
Again, in what follows, we decouple the antisymmetric matter by taking the limit  $m\rightarrow \infty$, but only after evaluating the integrals. The definition of the defect operator vev is then
\begin{align}
\label{YvevSp1}
&\left\langle \left[Y^{(k)}_{\pi}(M)\right]^{\pm 1} \right\rangle \equiv \frac{(-1)^k}{2}\left[\left\langle \left[Y^{(k,+)}(M)\right]^{\pm 1} \right\rangle - \left\langle \left[Y^{(k,-)}(M)\right]^{\pm 1} \right\rangle\right]\;\; ,\\
& \;\;\;\;\; \left\langle \left[Y_{\pi}(M)\right]^{\pm 1} \right\rangle=\sum_{k=0}^\infty \widetilde{\fq}^{k} \left\langle \left[Y^{(k)}_{\pi}(M)\right]^{\pm 1} \right\rangle \; .\nonumber
\end{align}
The vev $\left\langle Y_{\pi}(M)\right\rangle$ differs from the physical partition function $Z_{\pi}(M)$ only in the choice of contours, by avoiding all poles from the defect factor $Z^{(k,\pm)}_{defect}$.

Using a ``$i \E$" prescription or the Jeffrey-Kirwan residue, we find by direct computation that the partition function can be expressed as an infinite Laurent series in $Y$-operator loop vevs:
\begin{align}
\label{Sp1part}
Z^{inst}_{\pi}(M) = \left\langle Y_{\pi}(M) \right\rangle +   \ldots 
\end{align}
Indeed, the dots stand for an infinite number of $M$-dependent poles coming from $Z^{(k,\pm)}_{defect}$ that need to be enclosed by the contours, and each term in the $Y$-series stands for a residue at a new $M$-pole locus \cite{Haouzi:2020yxy}. Unlike $SU(N)$ gauge theories, there does not exist as of today a combinatorial interpretation of the $M$-poles in the index. It follows that there is no closed-form expression for the above expansion, and one needs to carry out the integrals explicitly to write down the higher order terms.\\

Nevertheless, let us press on and consider the normalized partition function, $Z^{inst}_{\pi}(M)/Z^{inst}_{\text{pure}, \pi}$, where the denominator is the partition function in the absence of the Wilson loop, with components
\begin{align}
\label{5dintegralCNpure}
& Z^{(k,\pm)}_{\text{pure}}  = \frac{1}{|O(k)^{\pm}|^\chi} \, \oint \left[\frac{d\phi_I}{2\pi i}\right]Z^{(k,\pm)}_{vec}\cdot Z^{(k,\pm)}_{antisym}
\end{align} 
and 
\beq
\label{Sp1partitionpure}
Z^{(k)}_{\text{pure}, \pi} \equiv \frac{(-1)^k}{2}\left[Z^{(k,+)}_{\text{pure}, \pi} - Z^{(k,-)}_{\text{pure}, \pi}\right]\;\; , \;\;\;\;\; Z^{inst}_{\pi}=\sum_{k=0}^\infty \widetilde{\fq}^{k}\; Z^{(k)}_{\text{pure}, \pi} \; .
\eeq
We should worry whether or not this normalized index is counting unwanted states present in the UV complete quantum mechanics but absent in the low energy QFT.
In the language of section 2, there are spurious contributions $Z^{extra}$ in the defect partition function, and $Z^{extra}_{pure}$ in the pure partition function: 
\beq
\label{extranormalizedagain}
\frac{Z^{inst}(M)}{Z^{inst}_{\text{pure}}} = \frac{Z(M)}{\langle 1 \rangle} \cdot \frac{Z^{extra}(M)}{Z^{extra}_{\text{pure}}} \, .
\eeq
 In the $SU(2)_{\pi}$ formalism, we had both $Z^{extra}=1$ and $Z^{extra}_{pure}=1$, which made our task easy. In the $Sp(1)_{\pi}$ formalism, however, both factors are nontrivial, and they do not cancel out:  $Z^{extra}/Z^{extra}_{\text{pure}} \neq 1$. Thankfully, the origin of  $Z^{extra}$ and $Z^{extra}_{pure}$  is easy to identify in our string theory construction: they are  D0/D$4'$/O8 and D0/O8 contributions, respectively. Indeed, even in the absence of D4 branes, the index is counting D0 brane states from the above sectors. It follows that the spurious factors could be identified by turning off the D4 brane fugacity in the partition function; we will see in a few paragraphs that this intuition is correct, and the decoupling of $Z^{extra}$  amounts to setting the 5d Coulomb parameter $a$ to $-\infty$ in the unnormalized  partition function, after evaluation of the integrals\footnote{There is also a ``center of mass" term present at the first instanton order only. We  give explicit formulas below.}:
 \beq
 \label{extraSp1}
 Z^{inst}(M)\big|_{a=-\infty} \; .
 \eeq 
From equation \eqref{extranormalizedagain}, it is then an easy task to extract the QFT quantity $Z(M)/{\left\langle 1 \right\rangle}$. After doing so, we expand the normalized partition function in the defect fugacity $e^{M}$, and derive a Schwinger-Dyson identity:
\beq
\label{SchwingerSp1pure}
\boxed{\frac{1}{\left\langle 1 \right\rangle}\bigg[	\left\langle Y_{\pi}(M) \right\rangle + \ldots \bigg]  = e^{-M}\,\cT(M; \{\beta_i^{\epsilon_1, \epsilon_2}\})} \; .
\eeq
Quite beautifully, we find that the right-hand side $\cT(M; \{\beta_i^{\epsilon_1, \epsilon_2}\})$ is precisely the polynomial \eqref{SchwingerA1pure} we previously computed in the $SU(2)_{\pi}$ formalism, at least up to 4 instantons. We conjecture this remains true to all instantons orders. In particular, it is a \emph{finite} polynomial in $e^M$, whose roots $\{\beta_i^{\epsilon_1, \epsilon_2}\}$ define the vacuum. A general proof of this statement would require a closed-form expression for the left-hand side of the Schwinger-Dyson equation, as was the case in section 2, where the classification of $M$-poles was nearly trivial. Here, the left-hand side has an infinite number of terms, in one-to-one correspondence with the infinite number of $M$-poles to be enclosed in the integrand, with a seemingly unruly behavior. Therefore, the finitess of the polynomial $\cT(M; \{\beta_i^{\epsilon_1, \epsilon_2}\})$ remains a conjecture in the $Sp(1)_{\pi}$ formalism. The Jeffrey-Kirwan residue prescription provides an explicit algorithm to check this conjecture to arbitrarily high instanton number.\\

We find it convenient to switch to exponentiated fugacities:
\begin{align}
\label{5dvariablesagain}
&z\equiv e^M \;\; , \;\;\;\;\; \alpha\equiv e^a \;\; , \;\;\;\;\; q\equiv e^{\E_1} \;\; , \;\;\;\;\; t\equiv e^{-\E_2}\\
&v\equiv\sqrt{\frac{q}{t}}=e^{\E_+} \;\; , \;\;\;\;\; u\equiv \sqrt{q\, t}=e^{\E_-}
\end{align}
In those variables, the Schwinger-Dyson identity reads
\beq
\label{SchwingerSp1pure5d}
\frac{1}{\left\langle 1 \right\rangle}\bigg[	\left\langle Y_{\pi}(z) \right\rangle + \ldots \bigg]  = z^{-1}\,\cT(z; \{\beta_i^{q,t}\}) \; .
\eeq
Assuming our conjecture on the regularity of the partition function in $z$ holds to all instanton orders, we turn to the interpretation of the result.

\vspace{10mm}

------\; {\bf Physics of the Schwinger-Dyson Identity}\; ------\\

First, can we make sense of equation \eqref{SchwingerSp1pure5d} as a quantization of the $Sp(1)_{\pi}$ Seiberg-Witten curve? This is a priori puzzling, because the left-hand side of the identity contains an infinite number of terms, whereas the Seiberg-Witten curve of $Sp(1)$  only contains $Y$-terms up to order  $\widetilde{\fq}^2$ in our conventions. We propose that the resolution of this apparent paradox is the same as in the  $\theta=0$ case \cite{Haouzi:2020yxy}. Namely, in the flat space limit $q, t \rightarrow 1$, each term starting at order $\widetilde{\fq}^3$ on the left-hand side of the Schwinger-Dyson identity must vanish, while the lower order terms must survive. Meanwhile, the Laurent polynomial on the  right-hand side becomes  $z^{-1}\,\cT(z; \{\beta_i^{1,1}\})$, where the roots $\{\beta_i^{1,1}\}$ are simply the flat space limit of the roots $\{\beta_i^{q, t}\}$. The exact mechanism of these cancellations deserves a careful investigation, and we leave it to future work.\\

Second, since the left-hand side of \eqref{SchwingerSp1pure5d} is regular in $z$, it may still be possible to interpret it as a $qq$-character of some representation of a quantum affine algebra. A hint should come from the study of quantum integrable systems with open boundaries, where the role of our orientifold plane in string theory is played by a reflection matrix for a XXZ spin chain \cite{Cherednik:1985vs,Sklyanin:1988yz}.\\

Third, just like in the $SU(2)$ formalism, the $z$-independent term on the right-hand side of the Schwinger-Dyson identity should be a $Sp(1)_{\pi}$ Wilson loop vev, where the trace of the loop is evaluated in the fundamental representation:
\beq
\label{RHSwilsonagain}
z^{-1}\,\cT(z; \{\beta_i^{q, t}\})= z - \langle W_{\bold{2}} \rangle + z^{-1} \; .
\eeq
As we mentioned earlier, we find by direct computation that the Wilson loop vev agrees exactly with the $SU(2)_{\pi}$ one computed in \eqref{Wilsonexplicit}, at least up to 4 instantons:
\beq
\label{Wilsonexplicitagain}
\langle W_{\bold{2}} \rangle = \alpha+\alpha^{-1} +\widetilde{\fq}\,\frac{\alpha^2\,(v+v^{-1})}{(1-(v\,\alpha)^2)(1-(\alpha/v)^2)} +o\left[\widetilde{\fq}^2\right]
\eeq

For the sake of completeness, we also give the UV spurious contribution of the D0/D$4'$/O8 system, which should be subtracted from the normalized ADHM index in order to obtain all the above results\footnote{In this additive notation,
	\beq
	\frac{Z(M)}{\left\langle 1 \right\rangle}= \frac{Z^{inst}(M)}{Z^{inst}_{\text{pure}}} - Z^{subtract}(M)
	\eeq 
	Equivalently, in the multiplicative notation of equation \eqref{extranormalizedagain}, 
	\beq
	\frac{Z^{extra}(M)}{Z^{extra}_{\text{pure}}}= \frac{Z^{inst}(M)/Z^{inst}_{\text{pure}}}{Z^{inst}(M)/Z^{inst}_{\text{pure}}- Z^{subtract}(M)}
\eeq}:
\begin{align}
\label{extraexplicitpi}
 Z_{\pi}^{subtract}(M) &=\frac{\widetilde{\fq}\,\sqrt{q\, t}}{(1-q)(1-t)} - Z^{inst}_{\pi}(M)\bigg|_{\alpha=0}\\
  &= \widetilde{\fq}\;\left(\frac{\sqrt{q\, t}}{(1-q)(1-t)} - \sum_{i=1}^\infty \chi^{SU(2)}_{2i}(v)\,z^{2i}\right)
\end{align}
Above, $\chi^{SU(2)}_{j}(v)$ is the character of the j-dimensional irreducible representation of $SU(2)$; for example, $\chi^{SU(2)}_{2}(v)=\sqrt{q/t}+\sqrt{t/q}$.\\

For comparison, when $\theta=0$, the spurious UV contribution is computed to be
\begin{align}
\label{extraexplicitzero}
 Z_{0}^{subtract}(M) &= - Z^{inst}_{0}(M)\bigg|_{\alpha=0}\\
&= -\widetilde{\fq}\; \sum_{i=1}^\infty \chi^{SU(2)}_{2i-1}(v)\,z^{2i-1}
\end{align}
We see that the spurious contribution when $\theta=0\; (\pi)$  is a generating function of odd (even) dimensional $SU(2)$ characters, and only affects the one instanton result.

\section{Generalizations}

In this section, we generalize our results to higher rank gauge theories and discuss the addition of fundamental matter.

\subsection{$Sp(N)$ at $\theta$-angle $\pi$ for $N>1$}

Our $Sp(1)_{\pi}$ analysis generalizes at once to higher rank theories, by increasing the number of D4 branes. The partition function is computed exactly as in the previous section,
\beq
\label{SpNpartition}
Z^{(k)}_{\pi}(M) \equiv \frac{(-1)^k}{2}\left[Z^{(k,+)}(M) - Z^{(k,-)}(M)\right]\;\; , \;\;\;\;\; Z^{inst}_{\pi}(M)=\sum_{k=0}^\infty \widetilde{\fq}^{k}\, Z^{(k)}_{\pi}(M) ,
\eeq
with the  explicit 1-loop determinants written in the appendix. The non-perturbative Schwinger-Dyson equation takes the same functional form as in the rank 1 case, 
\beq
\label{SchwingerSpNpure}
\boxed{\frac{1}{\left\langle 1 \right\rangle}\bigg[	\left\langle Y_{\pi}(M) \right\rangle + \ldots \bigg]  = e^{-N\, M}\,\cT(M; \{\beta_i^{\epsilon_1, \epsilon_2}\})} \; .
\eeq
This happens because the infinite Laurent series in $Y$-operator vevs on the left-hand side is entirely determined by the poles depending on $M$. Since we have not increased the number of D$4'$ branes, the $M$-dependent poles are the same for all $Sp(N)$ gauge theories. In particular, the associated $qq$-character has the same functional expression for all $N$.

The dependence on the rank is only present in the right-hand side of the Schwinger-Dyson identity, where $\cT(M; \{\beta_i^{\epsilon_1, \epsilon_2}\})$ is a finite polynomial of degree $2 N$  in $e^M$. As before, the roots $\{\beta_i^{\epsilon_1, \epsilon_2}\}$ define the vacuum.  We suspect that the coefficients of the polynomial can be reinterpreted as Wilson loop vevs of $Sp(N)$, albeit in a rather nontrivial way compared to the rank 1 case. To see how this works, let us consider a setup without any D0 brane. Then, the partition function is entirely due to the contribution of D4/D$4'$ strings,
\ie
Z_{D4/D4'}=\prod_{i=1}^N\sh(M\pm a_i) \; .
\fe
It is always possible to rewrite the above in terms of Wilson loop vevs, at least classically. For instance, when $N=2$, we obtain
\ie
Z_{D4/D4'}= z^{-2} -  \langle W_{\bold{[1,0]}} \rangle \,  z^{-1} + \left( \langle W_{\bold{[0,1]}} + 1 \rangle \right) -  \langle W_{\bold{[1,0]}} \rangle \, z + z^2\; ,
\fe
where we denoted the fundamental representations of $Sp(2)$ in Dynkin basis. After adding the D0 branes, the coefficients will get an infinite number of instanton corrections. They  are computed as usual by expanding the normalized partition function in the defect fugacity $z$. We leave the computation of these exact $Sp(N)$ Wilson loop vevs to future work.\\

We should address the presence of extra UV degrees of freedom in the ADHM index, which as usual need to be decoupled to derive the QFT Schwinger-Dyson identity. Thankfully, the increasing number of D4 branes does not create extra complications, and we find that the spurious factors can still be identified as in the $Sp(1)$ case, by decoupling the D4 branes. For instance, for $Sp(2)_{\pi}$, we find
\begin{align}
\label{extraexplicitpiSp2}
Z_{\pi}^{subtract}(M) &= - Z^{inst}_{\pi}(M)\bigg|_{\alpha_1, \alpha_2=0}\\
&= \widetilde{\fq}\; \sum_{i=1}^\infty \chi^{SU(2)}_{2i-1}(v)\,z^{2i-1}
\end{align}

As a last remark, after taking the flat space limit $\E_+, \E_- \rightarrow 0$, \eqref{SchwingerSpNpure} should become the Seiberg-Witten curve of the $Sp(N)_{\pi}$ theory after an infinite number of cancellations on the left-hand side, following the same reasoning as in the rank 1 case.

\vspace{10mm}

------\; {\bf Adding fundamental matter}\; ------\\

$N_f$ fundamental hypermultiplets contribute additional fermionic zero-modes to the ADHM index, and can be realized in the string theory by adding $N_f$ D8 branes in the $X_{0,1,2,3,4,5,6,7,8}$-directions. The relevant 1-loop determinants are written in the appendix.

The resulting Schwinger-Dyson equations will be slightly modified compared to the one we studied so far, but only barely: the $qq$-character will get an additional twist due to flavor factors, but the functional dependence on $Y$-operators will be the same as before, since there are no new $M$-poles to consider. Details in the case $\theta=0$ are given in \cite{Haouzi:2020yxy}.\\ 

If we require the 5d $Sp(N)$ gauge theory to admit a nontrivial UV fixed point, there are additional constraints on the theory; for instance, for a fixed rank $N$, only $N_f \leq 2 N + 4$ flavors are allowed \cite{Intriligator:1997pq}\footnote{The case $N=1$ is special and allows $N_f\leq 7$.}. When the inequality is saturated, the ADHM quantum mechanics sees a new continuum opening up in its Coulomb branch, which will manifest itself as a new UV contribution $Z^{extra}$ to the index. This will come from the D0/D$4'$/O8/D8 sector of the string theory, and decoupling such a spurious factor may require extra care.

\subsection{$SU(N)$ at Chern-Simons level $\kappa$ for $N>2$}

When $N>2$, the 5d $\cN=1$ $SU(N)$ theory admits an honest Chern-Simons term. The formalism we used for the theta angle of $SU(2)$ is readily generalized to study the Wilson loop of a higher rank special unitary theory, at Chern-Simons level $\kappa\in\{0,1,2,\ldots,N-1\}$. It is once again convenient to use a $(p,q)$ web of branes for the underlying UV picture. As in the $N=2$ case, we compute the partition function of a $U(N)$ theory first, by introducing $N$ moduli $a_{1,2,\ldots,N}$, and then freeze an overall $U(1)$ by imposing the traceless condition $\sum_{i=1}^N a_i =0$ to obtain the $SU(N)$ Coulomb moduli. The index is given by
\begin{align}
\label{5dintegralA1more}
& Z^{inst}(M)  =\sum_{k=0}^{\infty}\;\frac{\widetilde{\fq}^{k}}{k!} \, \oint \left[\frac{d\phi_I}{2\pi i}\right]Z^{(k)}_{vec}\cdot Z^{(k)}_{C.S.}\cdot  Z^{(k)}_{defect}(M)  \; , \\
&Z^{(k)}_{vec} =\prod_{\substack{I\neq J\\ I,J=1}}^k\sh(\phi_I-\phi_J)\prod_{I, J=1}^k\frac{\sh(\phi_I-\phi_J+2\E_+)}{\sh(\phi_I-\phi_J+\E_+\pm\E_-)}\prod_{I=1}^k\prod_{i=1}^N\frac{1}{\sh(\pm(\phi_I-a_i)+\E_+)}\; ,\\
&Z^{(k)}_{C.S.} =(-1)^{k\, \kappa}\prod_{I=1}^k e^{\phi_I\, \kappa}\; ,\\
&Z^{(k)}_{defect} =\prod_{i=1}^N\sh(a_i-M) \prod_{I=1}^k\frac{\sh(\pm(\phi_I-M)+\E_-)}{\sh(\pm(\phi_I-M)-\E_+)}\, .
\end{align}
Compared to the $N=2$ case, there are many more poles to be picked up by the contours\footnote{The additional poles to be picked up have a famous classification in terms of $N$-colored Young diagrams \cite{Nekrasov:2003rj}.}. However, from the Schwinger-Dyson perspective, the only poles to keep track of are the ones depending on the defect fugacity $M$. Then, it suffices to note that the contours of the $SU(N)$ partition function still enclose a pole at
\begin{equation}
\label{onlyonepoleUN}
\phi_I-M-\epsilon_+=0 \;\;\mbox{for some} \; I\in\{1,\ldots, k\} \; ,
\end{equation}  
and only there, for each instanton number $k$. We therefore conclude that the partition function can be written as
\beq
\label{partitionA1purekappa}
Z^{inst}(M)=	\left\langle Y(M) \right\rangle + \widetilde{\fq} \;(-1)^\kappa\, e^{\kappa(M+\E_+)} \left\langle\frac{1}{Y(M+2\,\epsilon_+)}\right\rangle\; .
\eeq 
Expanding the normalized partition function it in the defect fugacity $e^M$, we derive the  following Schwinger-Dyson identity for $SU(N)$ at level $\kappa$:
\beq
\label{SchwingerA1purekappa}
\boxed{\frac{1}{\left\langle 1 \right\rangle}\left[	\left\langle Y(M) \right\rangle + \widetilde{\fq} \;(-1)^\kappa\, e^{\kappa(M+\E_+)} \left\langle\frac{1}{Y(M+2\,\epsilon_+)}\right\rangle\right]  = e^{-M N/2}\,\cT(M; \{\beta_i^{\epsilon_1, \epsilon_2}\})} \; .
\eeq 
This is understood as a quantization of the Seiberg-Witten geometry, and the usual Seiberg-Witten curve of 5d $SU(N)$ at level $\kappa$ is recovered in the flat space limit $\E_+, \E_- \rightarrow 0$. On the right-hand side, $\cT(M; \{\beta_i^{\epsilon_1, \epsilon_2}\})$ is a finite polynomial of degree $N$ in $e^M$, whose roots $\{\beta_i^{\epsilon_1, \epsilon_2}\}$ define the vacuum.  In exponentiated variables, 
\begin{align}
\label{5dvariablesagainagain}
&z\equiv e^M \;\; , \;\;\;\;\; \alpha\equiv e^a \;\; , \;\;\;\;\; q\equiv e^{\E_1} \;\; , \;\;\;\;\; t\equiv e^{-\E_2}\\
&v\equiv\sqrt{\frac{q}{t}}=e^{\E_+} \;\; , \;\;\;\;\; u\equiv \sqrt{q\, t}=e^{\E_-}
\end{align}
the Schwinger-Dyson identity reads
\beq
\label{Schwinger5dkappa}
\frac{1}{\left\langle 1 \right\rangle}\left[\left\langle Y(z) \right\rangle + \widetilde{\fq} \;(-1)^\kappa\, z^\kappa \, v^\kappa\, \left\langle\frac{1}{Y(z\,v^2)}\right\rangle\right]  = z^{-1}\, \cT(z; \{\beta_i^{q,t}\})\; ,
\eeq
The absence of $z$-poles in $\cT(z; \{\beta_i^{q,t}\})$ is proved as in the $N=2$ case, by showing that one cannot have pinch singularities. 

The regularity in $z$ is equivalent to the statement that the left-hand side of the Dyson-Schwinger identity defines a $qq$-character for the spin 1/2 representation of the quantum affine algebra $U_q(\widehat{A_1})$. The Chern-Simons level manifests itself as an additional twist in the second term of the character, 
\beq
\label{qqcharacterkappa}
\chi^{(A_1, \kappa)}_{qq}=\frac{1}{\left\langle 1 \right\rangle}\left[\left\langle Y(z) \right\rangle + \widetilde{\fq} \;(-1)^\kappa\, z^\kappa \, v^\kappa\, \left\langle\frac{1}{Y(z\,v^2)}\right\rangle\right] \; ,
\eeq
This character can be built out of the following i-Weyl group action on the highest weight $\left\langle Y(z) \right\rangle$:
\beq
\label{iWeylkappa}
\left\langle Y(z) \right\rangle  \mapsto   \widetilde{\fq} \;(-1)^\kappa\, z^\kappa \, v^\kappa\, \left\langle\frac{1}{Y(z\,v^2)}\right\rangle \; .
\eeq
Meanwhile, the normalized index is the generating function of $SU(N)$ Wilson loops in all fundamental representations of the gauge group, as can be checked explicitly by carrying out an expansion in $z$.\\

In the case $\kappa=N$, a continuum appears in the 1d Coulomb branch, which manifests itself as a nontrivial $Z^{extra}$ contribution to the index, and is an artifact of the UV completion. In the IIB $(p,q)$ web picture, this is due to the presence of parallel NS5 branes, allowing a D1 brane continuum to appear. We already commented on that effect in the $N=2$ case, see figure \ref{fig:E1level2}. The decoupling of these unwanted states needs additional care in general.

\vspace{10mm}

------\; {\bf Adding fundamental matter}\; ------\\
	
We can add $N_f$ fundamental hypermultiplets by having $N_f$ semi-infinite D5 branes in the $(p,q)$ web. This results in a new D1/D5 sector, with new fermionic zero modes coming from Fermi multiplets in the bifundamental representation of $U(k)\times U(N_f)$. We therefore need to modify the index to account for these new Fermi multiplets in the quantum mechanics:
\begin{align}
\label{5dintegralA1fund}
& Z^{inst}(M)  =\sum_{k=0}^{\infty}\;\frac{\widetilde{\fq}^{k}}{k!} \, \oint_{\mathcal{M}_k}  \left[\frac{d\phi_I}{2\pi i}\right]Z^{(k)}_{vec}\cdot Z^{(k)}_{fund}\cdot  Z^{(k)}_{defect}  \; , \\
&Z^{(k)}_{fund} =(-1)^{k\, N_f/2}\prod_{I=1}^{k} \prod_{d=1}^{N_f} \sh\left(\phi_I-m_d\right) \; ,\label{fundQA1}
\end{align}
If $N_f$ is even (odd), then $\kappa$ is integer (half-integer). The phase  $(-1)^{k\, N_f/2}$ is introduced by hand to combine with the phase $(-1)^{k\, \kappa}$ of the bare Chern-Simons term. This results in an effective Chern-Simons level $\kappa+N_f/2$, consistent with field theory expectations in the presence of $N_f$ fundamental hypermultiplets. A first-principle derivation of this phase is not known to this day.\\ 

The matter factor $Z^{(k)}_{fund}$ does not contribute new $M$-poles, so the partition function will have the same functional form as previously. We find
\begin{align}
\label{A1fund}
Z^{inst}(M) = \left\langle Y(M) \right\rangle + \widetilde{\fq}\; (-1)^{\kappa+N_f/2}\, e^{\kappa(M+\E_+)} \prod_{d=1}^{N_f} \sh\left(M+\epsilon_+-m_d\right)  \left\langle\frac{1}{Y(M+2 \, \epsilon_+)}\right\rangle\; .
\end{align}

Expanding the normalized index in the fugacity $e^M$, we derive the following Schwinger-Dyson identity:
\beq
\label{SchwingerA1purekappafund}
\resizebox{.98\hsize}{!}{$\boxed{\frac{1}{\left\langle 1 \right\rangle}\left[	\left\langle Y(M) \right\rangle + \widetilde{\fq} \;(-1)^{\kappa+N_f/2}\, e^{\kappa(M+\E_+)} \, \prod_{d=1}^{N_f} \sh\left(M+\epsilon_+-m_d\right) \left\langle\frac{1}{Y(M+2\,\epsilon_+)}\right\rangle\right]  = e^{-M N/2}\,\cT(M; \{\beta_i^{\epsilon_1, \epsilon_2}\})} \; .$}
\eeq 
This is the quantized Seiberg-Witten geometry of 5d $SU(N)$ at Chern-Simons level $\kappa$ and with $N_f$ flavors, and the left-hand side is the associated $qq$-character. The usual Seiberg-Witten curve is recovered in the flat space limit $\E_+, \E_- \rightarrow 0$.\\

If we denote the flavor masses as $f_d\equiv e^{m_d}$, we can rewrite the identity in exponential variables, as
\beq
\label{Schwinger5dkappafund}
\frac{1}{\left\langle 1 \right\rangle}\left[\left\langle Y(z) \right\rangle + \widetilde{\fq} \;(-1)^{\kappa+N_f/2}\, z^\kappa \, v^\kappa\,\prod_{d=1}^{N_f} \left(\left(\frac{z\,v}{f_d}\right)^{1/2}- \left(\frac{f_d}{z\,v}\right)^{1/2}\right) \left\langle\frac{1}{Y(z\,v^2)}\right\rangle\right]  = z^{-\frac{N}{2}}\, \cT(z; \{\beta_i^{q,t}\}).
\eeq
The absence of $z$-poles in $\cT(z; \{\beta_i^{q,t}\})$ guarantees that the left-hand side defines a $qq$-character, with an additional twist due to the $N_f$ flavors. We will rewrite it slightly differently, to make explicit the dependence on the effective Chern-Simons level $\kappa+N_f/2$:
\beq
\label{qqcharacterkappaflavors}
\chi^{(A_1, \kappa, N_f)}_{qq}=\frac{1}{\left\langle 1 \right\rangle}\left[\left\langle Y(z) \right\rangle + \widetilde{\fq} \;\frac{(-z\, v)^{\kappa+N_f/2}}{\prod_{d=1}^{N_f} f_d^{1/2}}\,\prod_{d=1}^{N_f} \left(1- \frac{f_d}{z\,v}\right) \left\langle\frac{1}{Y(z\,v^2)}\right\rangle\right] \; .
\eeq

For a fixed rank $N$ and Chern-Simons level $\kappa$, the gauge theory is known to have a UV fixed point when $N_f \leq 2 N - 2 \kappa$. When the inequality is saturated, a continuum appears in the Coulomb branch of the ADHM quantum mechanics, and a spurious contribution $Z^{extra}$ needs to be factored out of the index. This can once again be detected from parallel NS5 branes in the $(p,q)$ web picture.\\

As a last remark, more exotic cases can be considered. For instance, when $N\leq 8$, there can be 1 antisymmetric and $N_f$ fundamental hypermultiplets, and a UV fixed point exists when $N_f \leq 8-N - 2 \kappa$. An ADHM description for antisymmetric matter is known \cite{Shadchin:2004yx}, and non-perturbative Schwinger Dyson equations have been discussed in the case $\kappa=0$ \cite{Haouzi:2020yxy}. It should be possible to generalize the analysis to an arbitrary Chern-Simons level following the methods we presented here.

\vspace{20mm}

\section*{Acknowledgments}
We thank Ori Ganor, Jihwan Oh, Vivek Saxena and Luigi Tizzano for valuable discussions at early stages of this project. The research of NH is supported by the Simons Center for Geometry and Physics.

\pagebreak

\begin{appendix}
	
	\section{1-loop determinants for $G=Sp(N)$ and $\widehat{G}=O(k)$}
	
	We use the notation $\sh(x)\equiv 2 \sinh(x/2)$ and $\ch(x)\equiv 2 \cosh(x/2)$.  Products over all signs inside an argument have to be considered, and we introduce the integer $\chi=0, 1$. 
	
	For a given instanton number $k$, the index receives the following contribution, written in the notation of the main text as
	\begin{align}
	\label{5dintegralCNagain}
	& Z^{(k,\pm)}(M)  = \frac{1}{|O(k)^{\pm}|^\chi} \, \oint \left[\frac{d\phi_I}{2\pi i}\right]Z^{(k,\pm)}_{vec}\cdot Z^{(k,\pm)}_{antisym}\cdot  Z^{(k,\pm)}_{defect}(M) \; .
	\end{align}
	
In this appendix, we will rewrite the above factors in terms of string theory quantities as follows:
\begin{align}
Z^{(k,\pm)}_{vec}\cdot Z^{(k,\pm)}_{antisym} &\equiv {Z}^{(k,\pm)}_{D0/D0}\cdot {Z}^{(k,\pm)}_{D0/D4}\nonumber\\ 
 Z^{(k,\pm)}_{defect}(M) &\equiv  {Z}^{(k,\pm)}_{D0/D4'}\cdot {Z}_{D4/D4'} 
\end{align}

The 1-loop determinants of the D0/D0 strings are given by
	\ie\label{Zplus}
	\hspace*{-1cm}{Z}_{D0/D0}^{(k=2n+\chi, +)}=& \bigg[\Big(\prod_{I=1}^{n} \sh(\pm \phi_I)\Big)^{\chi }\prod_{I<J}^{n} \sh(\pm \phi_I \pm \phi_J) \bigg]\sh(2\epsilon_+)^n \Big(\prod_{I=1}^{n}\sh(\pm \phi_I + 2\epsilon_+)\Big)^{\chi}\prod_{I < J}^{n}\sh(\pm \phi_{I} \pm \phi_{J} + 2\epsilon_+)
	\\
	&\hspace*{-.5cm}\times\sh(\pm m - \epsilon_-)^n \Big(\prod_{I=1}^{n}\sh(\pm\phi_I \pm m - \epsilon_-)\Big)^\chi \prod_{I<J}^{n}\sh(\pm\phi_I \pm \phi_J \pm m - \epsilon_-) 
	\\
	&\hspace*{-.5cm}\times {1\over \sh(\pm m - \epsilon_+)^{n+\chi}}\Big(\prod_{I=1}^{n} \frac{1}{\sh(\pm\phi_I \pm m - \epsilon_+)}\Big)^{\chi}\prod_{I=1}^{n} \frac{1}{\sh(\pm 2\phi_I \pm m - \epsilon_+) } \prod_{I<J}^{n} \frac{1}{\sh(\pm\phi_I \pm \phi_J \pm m - \epsilon_+)}
	\\
	&\hspace*{-.5cm}\times{1\over \sh(\pm \epsilon_- + \epsilon_+)^{n+\chi}} \Big( \prod_{I=1}^{n} \frac{ 1} {\sh(\pm \phi_I \pm \epsilon_- + \epsilon_+)}\Big)^{\chi}
	\prod_{I=1}^{n} \frac{1}{\sh(\pm 2\phi_{I} \pm \epsilon_- + \epsilon_+)} \prod_{I < J}^{n} \frac{1 }{\sh(\pm \phi_{I} \pm \phi_{J} \pm \epsilon_- + \epsilon_+)},
	\fe
	and
	\ie\label{Zminusodd}
	\hspace*{-1cm}{Z}_{D0/D0}^{(k=2n+1, -)}=&\Big( \prod_{I=1}^{n} \ch(\pm \phi_I) \prod_{I<J}^{n} \sh(\pm \phi_I \pm \phi_J)\Big) \sh(2\epsilon_+)^n\prod_{I=1}^{n} \ch(\pm \phi_I + 2\epsilon_+)\prod_{I < J}^{n}\sh(\pm \phi_{I} \pm \phi_{J} + 2\epsilon_+)
	\\
	&\hspace*{-0.5cm}\times\sh(\pm m - \epsilon_-)^n \prod_{I=1}^{n}\ch(\pm\phi_I \pm m - \epsilon_-) \prod_{I < J}^{n}\sh(\pm\phi_I \pm \phi_J \pm m - \epsilon_-)
	\\
	&\hspace*{-0.5cm}\times\frac{1}{\sh(\pm m - \epsilon_+)^{n+1}}  \prod_{I=1}^{n} \frac{1}{\ch(\pm\phi_I \pm m - \epsilon_+)\sh(\pm 2\phi_I \pm m - \epsilon_+) }  \prod_{I<J}^{n} \frac{1}{\sh(\pm\phi_I \pm \phi_J \pm m - \epsilon_+)}
	\\
	&\hspace*{-0.5cm}\times \frac{1}{\sh(\pm \epsilon_- + \epsilon_+)^{n+1} }  \prod_{I=1}^{n} \frac{ 1} {\ch(\pm \phi_I \pm \epsilon_- + \epsilon_+)  \sh(\pm 2\phi_{I} \pm \epsilon_- + \epsilon_+)} \prod_{I < J}^{n} \frac{ 1}{\sh(\pm \phi_{I} \pm \phi_{J} \pm \epsilon_- + \epsilon_+)},
	\fe
	and
	\ie\label{Zminuseven}
	\hspace*{-2cm}{Z}_{D0/D0}^{(k=2n, -)}=&\Big(\prod_{I<J}^{n-1} \sh(\pm \phi_I \pm \phi_J) \prod_{I=1}^{n-1}\sh(\pm2 \phi_I) \Big) \ch(2\epsilon_+)(\sh{\epsilon_+} )^{n-1}\prod_{I=1}^{n-1} \sh(\pm 2\phi_I + 4\epsilon_+)   \prod_{I < J}^{n-1}  \sh(\pm \phi_{I} \pm \phi_{J} + 2\epsilon_+)
	\\
	&\hspace*{-0.5cm}\times \ch(\pm m - \epsilon_-)\, \sh(\pm m - \epsilon_-)^{n-1}\prod_{I=1}^{n-1}\sh(\pm 2\phi_I \pm 2 m -2 \epsilon_-) \prod_{I<J}^{n-1}\sh(\pm\phi_I \pm \phi_J \pm m - \epsilon_-)
	\\
	&\hspace*{-0.5cm}\times \frac{1}{\sh(\pm m - \epsilon_+)^n \sh(\pm 2 m -2 \epsilon_+)} \prod_{I=1}^{n-1} \frac{1}{\sh(\pm 2\phi_I \pm 2 m - 2 \epsilon_+)\sh(\pm 2\phi_I \pm m - \epsilon_+)} 
	\prod_{I<J}^{n-1} \frac{1}{\sh(\pm\phi_I \pm \phi_J \pm m - \epsilon_+)}
	\\
	&\hspace*{-0.5cm}\times \frac{1}{\sh(\pm \epsilon_- + \epsilon_+)^n \sh(\pm 2 \epsilon_- + 2\epsilon_+)}  \prod_{I=1}^{n-1} \frac{1 }{\sh(\pm 2\phi_I \pm 2 \epsilon_- + 2 \epsilon_+) \sh(\pm 2\phi_{I} \pm \epsilon_- + \epsilon_+)}\prod_{I < J}^{n-1} \frac{1}{\sh(\pm \phi_{I} \pm \phi_{J} \pm \epsilon_- + \epsilon_+)}.
	\fe
	The first to the forth lines of the equations \eqref{Zplus}, \eqref{Zminusodd} and\eqref{Zminuseven}, are the 1-loop determinants of the $(0,4)$ vector multiplet, Fermi multiplet, twisted hypermultiplet and hypermultiplet, respectively.  The 1-loop determinants of the D0/D4 strings are given by
	\ie\label{D0-D4Integrand}
	&{Z}^{(k=2n+\chi, +)}_{D0/D4}=\prod_{i=1}^N\Big(\frac{\sh(m\pm a_i)}{\sh(\pm a_i+\epsilon_+)}\Big)^\chi\prod^n_{I=1}\frac{\sh(\pm\phi_I\pm a_i-m)}{\sh(\pm\phi_I\pm a_i+\epsilon_+)},
	\\
	&{Z}^{(k=2n+1, -)}_{D0/D4}=\prod_{i=1}^N\frac{\ch(m\pm a_i)}{\ch(\pm a_i+\epsilon_+)}\prod^{n}_{I=1}\frac{\sh(\pm\phi_I\pm a_i-m)}{\sh(\pm\phi_I\pm a_i+\epsilon_+)},
	\\
	&{Z}^{(k=2n, -)}_{D0/D4}
	%=Z^{-,\,k=2n+1}_{\text{\tiny D0-D4}}\Big|_{\phi_n=2\pi i}
	=\prod_{i=1}^N\frac{\sh(2m\pm2 a_i)}{\sh(\pm 2 a_i+2\epsilon_+)}\prod^{n-1}_{I=1}\frac{\sh(\pm\phi_I\pm a_i-m)}{\sh(\pm\phi_I\pm a_i+\epsilon_+)}.
	\fe
	In order to deduce the contribution of D0/D$4'$ strings, we make use of a symmetry of the brane setup \ref{table:BranesSpN}. Namely, under the exchange of the coordinates $X_{1,2,3,4}  \leftrightarrow X_{5,6,7,8}$, D4 and D$4'$ branes get swapped, while the D0 branes and the O$8^-$ orientifold plane are invariant.  Hence,  we can write the D0/D$4'$ contribution from the D0/D4 one, after simply exchanging $\E_1,\E_2$ with $\E_3,\E_4$. This translates to
	\ie
	a_i\leftrightarrow M,\quad m\leftrightarrow \E_-,\quad \E_+\leftrightarrow -\E_+ \; ,
	\fe
	and we obtain
	\ie\label{D0-D4pIntegrand}
	&{Z}^{(k=2n+\chi, +)}_{D0/D4'}=\Big(\frac{\sh(\pm M-\E_-)}{\sh(\pm M-\epsilon_+)}\Big)^\chi\prod^n_{I=1}\frac{\sh(\pm\phi_I\pm M-\E_-)}{\sh(\pm\phi_I\pm M-\epsilon_+)},
	\\
	&{Z}^{(k=2n+1, -)}_{D0/D4'}=\frac{\ch(\pm M-\E_-)}{\ch(\pm M-\epsilon_+)}\prod^{n}_{I=1}\frac{\sh(\pm\phi_I\pm M-\E_-)}{\sh(\pm\phi_I\pm M-\epsilon_+)},
	\\
	&{Z}^{(k=2n, -)}_{D0/D4'}
	%=Z^{-,\,k=2n+1}_{\text{\tiny D0-D4}}\Big|_{\phi_n=2\pi i}
	=\frac{\sh(\pm 2 M- 2\E_-)}{\sh(\pm 2 M-2\epsilon_+)}\prod^{n-1}_{I=1}\frac{\sh(\pm\phi_I\pm M-\E_-)}{\sh(\pm\phi_I\pm M-\epsilon_+)}.
	\fe
	The 1-loop determinant of the D4/D$4'$ strings is responsible for a Fermi multiplet transforming in the bifundamental representation of $G\times G'= Sp(N)\times Sp(1)$. We obtain
	\ie
	Z_{D4/D4'}=\prod_{i=1}^N\sh(M\pm a_i) \; .
	\fe
	Fundamental matter is introduced through the introduction of D8 branes in the $X_{0,1,2,3,4,5,6,7,8}$ directions. The associated 1-loop determinants for the D0/D8 strings are
	\ie\label{D0-D8integrand}
	&{Z}^{(k=2n+\chi, +)}_{D0/D8}=\prod^{N_f}_{d=1}\sh(m_d)^\chi \prod^n_{I=1}\sh(\pm\phi_I + m_d),
	\\
	&{Z}^{(k=2n+1, -)}_{D0/D8}=\prod^{N_f}_{d=1}\ch(m_d)\prod^n_{I=1}\sh(\pm\phi_I + m_d),
	\\
	&{Z}^{(k=2n, -)}_{D0/D8}=\prod^{N_f}_{d=1}\sh(2 m_d) \prod^{n-1}_{I=1}\sh(\pm\phi_I + m_d).
	\fe
	Finally, the Weyl factors of the $O(k)_+$ and $O(k)_-$ components  are given by
	\begin{align}
	\hspace{-1cm}|O(k)^+|^{\chi=0} = \frac{1}{2^{n-1} n!} ,\;\; |O(k)^+|^{\chi=1} = \frac{1}{2^n n!} ,\;\; |O(k)^-|^{\chi=0} =  \frac{1}{2^{n-1}(n-1)!},\;\; |O(k)^-|^{\chi=1} = \frac{1}{2^n n!}.
	\end{align}

\end{appendix}

\newpage

\bibliography{thetaangleqq}
\bibliographystyle{JHEP}

\end{document}